\documentstyle[11pt,epsfig]{article}
 \def\baselinestretch{1.3}
\newcommand{\ba}{\begin{array}}
\newcommand{\ea}{\end{array}}
\newcommand{\bd}{\begin{displaymath}}
\newcommand{\ed}{\end{displaymath}}
\newcommand{\be}{\begin{equation}}
\newcommand{\ee}{\end{equation}}
\newcommand{\bea}{\begin{eqnarray}}
\newcommand{\eea}{\end{eqnarray}}

\def\etal{ {\em et al.}}

\def\q2 {q^2}

\def\lp{\lambda^{\prime}}

 \def\N10{\widetilde \chi_1^0}
                         \def\C1p{\widetilde \chi_1^+}
                         \def\C1m{\widetilde \chi_1^-}
                         \def\C1pm{\widetilde \chi_1^\pm}

\def\sneu {\tilde \nu}

\def\beq{\begin{eqnarray}}

\def\enq{\end{eqnarray}}

\def\lsim{\:\raisebox{-0.5ex}{$\stackrel{\textstyle<}{\sim}$}\:}
\def\gsim{\:\raisebox{-0.5ex}{$\stackrel{\textstyle>}{\sim}$}\:}
\begin{document}
 \begin{flushright}
{\large CERN-TH/2001-036}\\
{\large IPPP/01/62}\\
{\large DCPT/01/24}\\
{\large December 2001}\\
{\large Revised \today}\\
\end{flushright}
\begin{flushleft}
\end{flushleft}
\begin{center}
{\Large\bf A study of single sneutrino production in association\\[0.25cm]
 with fermion pairs at polarised photon colliders}\\[15mm]
{\bf Dilip Kumar Ghosh$^{a,}$\footnote{dghosh@phys.ntu.edu.tw}
and
Stefano Moretti$^{b,}$\footnote{Stefano.Moretti@cern.ch}}\\[4mm]
{\em $^{a}$ Department of Physics, National Taiwan University \\
Taipei, TAIWAN 10617, Republic of China}\\[4mm]
{\em $^{b}$ CERN Theory Division, CH-1211 Geneva 23, Switzerland\\
{\rm and}\\
Institute for Particle Physics Phenomenology,
University of Durham, Durham DH1 3LE, UK}\\[30mm]
\end{center}
\begin{abstract}
We investigate single sneutrino production in the context
of R-parity-violating Supersymmetry at future $\gamma\gamma$ linear
colliders. The sneutrino
is produced in association with fermion pairs and it is shown
that its decays into
two further fermions will lead to a clean signal.  We also discuss 
possible Standard Model backgrounds and the effects of beam polarisation.
\end{abstract}

\vskip 1 true cm

\noindent


\newpage
\setcounter{footnote}{0}

\def\baselinestretch{1.8}
\section{Introduction}

Supersymmetry (SUSY) is currently the most attractive theoretical
framework describing
physics beyond the Standard Model (SM). Even the minimal 
extension of the SM incorporating SUSY (MSSM) predicts a zoo of new particles,
which have not yet been observed. One of the major areas of activity in 
high energy physics today and in the near future is to prove their existence.
If SUSY is realised at the electroweak (EW) scale, many of the 
superparticles should be discovered at  
next generation hadron colliders, such as Tevatron (Run II, $\sqrt 
{s_{p\bar p}}=2$ TeV) 
at FNAL and the Large Hadron Collider (LHC, $\sqrt{s_{pp}}=14$ TeV) at CERN.  
These machines, while having the chance of being the first to access the SUSY 
domain, are however hampered by the fact that a large QCD background and
the lack of knowledge of the initial centre-of-mass (CM) partonic energies
render difficult the task of determining sparticle properties (masses,
couplings, quantum numbers, etc.). An insight into this `SUSY spectrum' 
would in fact shed light on the yet unknown mechanism leading to SUSY-breaking.

In contrast, in $e^+e^-$ collisions, the QCD noise is under control
and the initial energies of the leptons are generally well known.
This has contributed in the recent years to the generation of
 a strong consensus
behind the option of building electron-positron Linear Colliders (LCs), 
operating in the energy range from 500 GeV to 3 TeV, as the accelerators 
most suited to inherit the legacy of the Run II and LHC era 
\cite{SUSY_search}. Such machines would not only provide the ideal environment 
for discovering the SUSY particles which could be missed out at 
the FNAL and CERN experiments, but 
would also allow for the precise determination of 
the mentioned SUSY spectrum. For example, mass measurements are aided
by the ability to perform threshold scans by varying the collider CM energy.
Furthermore, the spin properties of many SUSY particles can be accessed
by exploiting an efficient beam polarisation, a feature 
altogether missing at the Tevatron and the LHC. 

Another advantage of LCs is that they can easily be converted to run 
quite simply in $e^-e^-$ mode or even in $e\gamma$ and $\gamma\gamma$, the 
latter by using Compton back-scattering of laser 
photons against the electrons/positrons \cite{ginzburg,telnov}, all
such collisions taking place
with energy and luminosity comparable to those obtainable from the primary 
$e^+e^-$ design. Quite apart from SUSY \cite{ememSUSY}, 
it should be recalled that
electron-electron collisions would constitute a privileged window on, 
{\it e.g.},  models with extended Higgs sectors whereas those employing
photon beams would easily allow for, {\it e.g.}, 
the study of a plethora of QCD topics.

To come back to SUSY, it should be mentioned that there have been in the recent
years quite promising explorations of the physics potential of 
$\gamma \gamma $ LCs as a probe of the low energy dynamics of the theory
\cite{Photon_phys}. It is the intention of our study to further
dwell on this topic, by considering the scope of LCs in accessing some
R-parity-violating (RPV) signals of SUSY. 

\section{R-parity-violating Supersymmetry}

The construction of the most general Supersymmetric extension of the
SM leads to Baryon-$(\rm B)$- and Lepton-$(\rm L)$-number-violating 
operators in the superpotential 
\begin{equation}\label{SuperP}
W_{\not R} = \epsilon_{i} \lambda_{ijk} {\hat L}_i {\hat L}_j {\hat E}_k^c +
\lambda_{ijk}' {\hat L}_i {\hat Q}_j {\hat D}_k^c  
+ \epsilon_i {\hat L}_i {\hat H}_2 +
\lambda_{ijk}''{\hat U}_i^c {\hat D}_j^c {\hat D}_k^c. 
\end{equation}
Here, ${\hat H}_1$, ${\hat H}_2$ are the $SU(2)$ doublets Higgs superfields
which give rise to the masses of down-type and up-type quark superfields, 
respectively, $\hat L(\hat Q)$ denotes lepton(quark) doublet superfields, 
${\hat E}^c$, ${\hat D}^c$, ${\hat U}^c$ are the singlet lepton and quark 
superfields, $i, j, k$ are the generational indices and we have suppressed 
the $SU(2)$ and $SU(3)$ indices. The $\lambda_{ijk}$ are anti-symmetric in 
$i$ and $j$ while the $\lambda_{ijk}''$ are anti-symmetric in $j$ and $k$.
The first three terms in $W_{\not R}$ violate lepton number and the last 
term violates baryon number conservation. The simultaneous presence
of both B- and L-violating operators would induce
rapid proton decay which would contradict the strict experimental
bound of \cite{proton_dk}. In order to keep the proton lifetime within
the experimental limit, one needs to impose an additional symmetry beyond
the SM gauge symmetry, in order to force the unwanted B- and L-violating 
interactions to vanish.  In most cases, this can be achieved
by imposing a discrete symmetry, called R-parity \cite{fayet},
defined as ${\rm R} = (-1)^{\rm 3B+L+2S}$, where S is the
spin. This symmetry not only forbids rapid proton
decay \cite{weinberg} but also renders stable the lightest
supersymmetric particle (LSP).  

However, R-parity is quite an {\it ad hoc} assumption in nature, 
as there are no strong theoretical arguments to support it. Therefore, 
it is much justified to investigate the phenomenological consequences
of RPV SUSY.
Extensive studies have been carried out in order to look for direct 
as well as indirect
evidence of trilinear R-parity violation in different processes 
at various colliders as well as in order to 
put constraints on various RPV couplings \cite{rpv_review}.

\begin{figure}[hbt]
\centerline{\epsfig{file=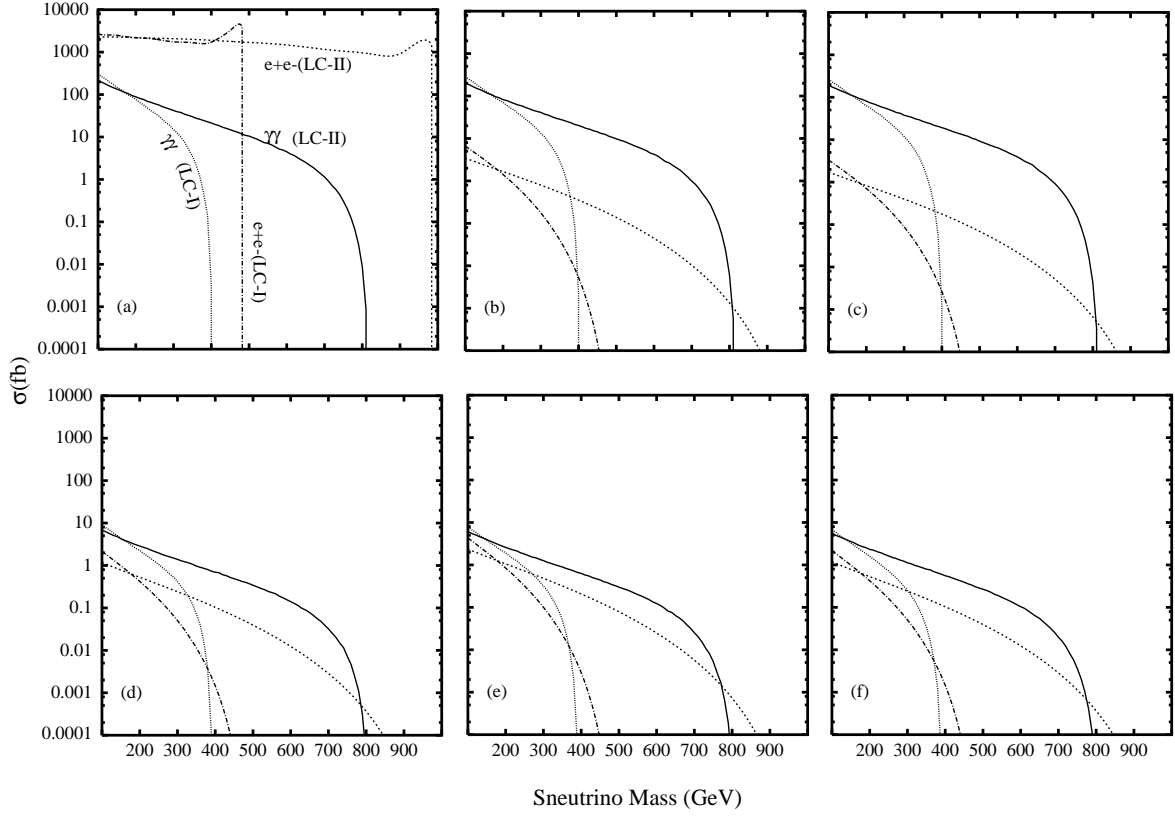,width=\linewidth}}
\vspace*{-6.0cm}
\caption{Cross sections for following production processes :
$\tilde \nu e^+e^- $ (a), 
$\tilde \nu \tau^+\mu^- + ( c.c.) $ (b), 
$\tilde \nu \tau^+\tau^- $ (c), 
$\tilde \nu s {\bar s} $ (d), 
$\tilde \nu b {\bar s} + (c.c.) $ (d), 
$\tilde \nu b {\bar b} $ (f), 
at $\gamma \gamma $ collider (solid) and their in $e^+e^-$ annihilation
(dashed), as a function of the sneutrino mass $M_{\sneu}$,
at $\sqrt{s_{e^+e^-}}=500$ GeV (LC-I) and 1 TeV (LC-II), after the cuts in
(\ref{cuts}). For simplicity, we have
set the parity violating couplings $\lambda$ and $\lambda^\prime$
to 1. }
\label{comparison}
\end{figure}

Resonant sneutrino production in 
$\gamma \gamma $ collisions has been studied 
in Ref.~\cite{Soni}, where the rare decays into two
photons or gluons were considered. 
In this article, we will consider instead RPV
single production of sneutrinos in association with fermion pairs 
in polarised photon-photon 
collisions at 500 GeV and 1 TeV LCs, and their subsequent decays into two 
further fermions, via trilinear L-violating operators, while
preserving B-conservation. The latter channel is in our opinion more suited as
a sneutrino `search' mode in $\gamma\gamma$ collisions than the former,
simply because one can scan a wider range of sneutrino masses ${M_{\tilde\nu}}$
(as long as $\sqrt{s_{\gamma\gamma}}\approx0.8\sqrt{s_{e^+e^-}}\gg M_{\tilde\nu}$),
thanks to the fact that some amount of energy is carried away by
the accompanying fermion pair, whereas in direct production the
only  $M_{\tilde\nu}$ attainable is basically the (reduced) CM energy itself.
Furthermore, the associate mode
may also induce flavour changing final states, so that -- as pointed 
out in \cite{sourov2} -- unlike in the case of resonant production,
one has that the corresponding signatures are basically
SM background free. 
Schematically, one has
\begin{equation}\label{production}
\gamma \gamma \to {\tilde \nu} \ell^\pm \ell^{\prime \mp}~~{\rm{or}}~~ 
{\tilde \nu} q {\bar q^\prime }
\end{equation}
{\rm{with}}
\begin{equation}\label{decay}
{\tilde \nu} \to \ell^{'' \pm} \ell^{''' \mp}~~{\rm{or}}~~ 
{\tilde \nu} \to q'' {\bar q}^{'''}, 
\end{equation}
where the $\ell$'s refer to $e,\mu$ and $\tau$ leptons and
the $q$'s to $d,u,s,c$ and $b$ quarks.
Finally, the main advantage of exploiting $\gamma\gamma$
collisions in place of $e^+e^-$ ones \cite{epem,Barger}
in producing single sneutrinos in association with a fermion pair
in final states of the type (\ref{production})
resides in the fact that the cross sections for the former
are generally larger than those for the latter, as one
can appreciate in Figure~\ref{comparison}. There, as an
illustration, we have plotted the unpolarised production rates
for both the $\gamma\gamma$ and $e^+e^-$ induced modes, using
the photon structure functions given in \cite{ginzburg}, at
$\sqrt{s_{e^+e^-}}=500$ GeV
 and 1 TeV. Apart from the $ {\tilde \nu} e^+e^-$ final state,
which in electron-positron annihilation 
receives very large additional contributions from small angle Bhabha-like 
scattering amplitudes (with respect to the other final states),
the photon processes are dominant over the electron-positron 
ones\footnote{We will defer the study of the $e^+e^-$ processes to another
paper \cite{preparation}. We should however mention here that we
have verified that,
given the final luminosities collected at LEP2 (see Ref.~\cite{LEP2lumi}),
the signatures considered in (\ref{production})--(\ref{decay})
but produced via $e^+e^-$ annihilations between 2060 and 210 GeV 
could have not been seen at the CERN machine, 
for the choice of RPV couplings adopted in the following
(see also \cite{snuLEP}).}.

The $\gamma\gamma$ induced associate production process
has been investigated recently in Ref.~\cite{sourov2}, by assuming 
unpolarised photon beams and without any detailed background estimates. 
We will improve on that study by exploiting
polarised $\gamma\gamma$ scatterings, 
as it has been shown
that a high degree of polarisation can be transmitted from the
electrons, positrons and laser photons to the 
Compton back-scattered photons, and by
including a study of the irreducible SM background\footnote{We make use
of HELAS \cite{HELAS} and MadGraph \cite{Tim} to produce the
helicity amplitudes, for both signal and backgrounds,
and integrate these numerically by using VEGAS \cite{VEGAS}.}. 
In fact, it will  be shown that polarisation may
help to improve the signal-to-background ratio ($S/B$) in some
instances.
We consider a general MSSM parameter space, with no assumption 
on the mechanism of SUSY-breaking, hence defining all parameters at
the EW scale.

Before proceeding to the analysis, it is
is useful to note at this point that the $\epsilon_i$ terms 
in (\ref{SuperP}) can in principle be removed by a 
re-definition of the lepton doublets ${\hat L}_i$,
which would in turn lead to their `absorption' into the 
$\lambda,\lp$ couplings and in the parameters of the scalar potential of 
the SUSY model. However, the $\epsilon_i$'s could then 
re-appear at a different energy 
scale. Bilinear terms could also lead to a possible vacuum expectation
value (VEV) for the sneutrino(s) and mixing of: ($a$) charged leptons with
charginos, ($b$) sleptons with charged Higgs bosons, ($c$) neutrinos with
neutralinos and ($d$) sneutrinos with neutral Higgs bosons.
This last mixing could indeed affect the process discussed here. 
However, this phenomenon is suppressed
by the small Yukawa couplings of our $\ell$ and $q$ fermions,
so that we feel justified in neglecting it here
({\it i.e.}, we are making the assumption that the $\epsilon_i$ terms are
small)\footnote{This would not be possible for processes involving
top (anti)quarks, because of their large mass. However,  in
(\ref{production})--(\ref{decay}), $t$ quarks contributions will 
have negligible impact, because strongly suppressed by phase space effects.
(Some phenomenological consequences of a sneutrino VEV and
L-violating mixing have been discussed in 
literature \cite{sourov1}.)}.

The paper is organised as follows. In section 3, we discuss the
phenomenology of processes (\ref{production})--(\ref{decay})  
in presence of polarised incoming photons.
In section 4 we present our numerical results
(including those for the backgrounds), followed 
by our conclusions in section 5.

\section{Singly produced sneutrinos at polarised photon colliders}

In the RPV MSSM, the sneutrino displays a coupling 
with pairs of
leptons ($\lambda$-type couplings) and quarks ($\lp$-type couplings). Single
production of sneutrino in association with fermion pairs in 
(\ref{production}) can occur through any of these
two types of L-violating couplings. Depending upon the nature of the 
vertex involved, the above process may also 
lead to flavour changing final states. 

The polarised photon flux and polarisation have been worked out 
in \cite{ginzburg} and are discussed in details 
in Ref. \cite{telnov}. For brevity,
we do not reproduce here those formulae, rather we simply 
recall to the un-familiar reader the basic features of
polarised $\gamma\gamma$ scatterings.

\begin{enumerate}
\item
We assume that the laser back-scattering parameter
 assumes its maximum value,   $z\equiv z_{\rm{max}} = 2(1 +
\sqrt{2}) \simeq 4.828$ \cite{ginzburg}. 
In fact, with increasing $z$ the high energy photon spectrum 
becomes more mono-chromatic. However,
for $z > z_{\rm{max}}$, the probability of $e^+e^-$ pair creation increases,
resulting in larger photon beam degradation.
\item
The reflected photon beam carries off only a fraction $x$ of the 
$e^\pm$ energy, with $x_{\rm{max}} = z/(1+z) \simeq 0.8$, while 
$x_{\rm{min}}= (M_{\tilde \nu} + m_{f} + m_{\bar f'})/\sqrt{s_{e^+e^-}}$
(hereafter, $f^{(')}=\ell,q$). 
\item
The polarization of the two initial laser ($\gamma$) and 
electron/positron ($e$) beams 
are defined by $P_{\gamma_-}, P_{\gamma_+}, P_{e^-}$ and $P_{e^+}$, 
respectively, where, for the first two quantities,
$-(+)$ identifies the laser colliding against
the electron(positron).
\item Finally, one can cast the polarised production cross-section
in the following form:
\begin{eqnarray} \nonumber
\sigma_{e^+e^-\to \gamma \gamma \to \sneu f \bar f^\prime} (s) 
=& \int &dx_- dx_+ F_-^{\gamma/e}(P_{e^-}, P_{\gamma_-}, x_-;P_-) 
                 F_+^{\gamma/e}(P_{e^+}, P_{\gamma_+}, x_+;P_+)\\ 
&\times&\hat\sigma_{\gamma \gamma \to \sneu f \bar f^\prime} (\hat s,P_-,P_+), 
\end{eqnarray}
where $x_{-(+)}$ is the electron(positron) momentum fraction carried
by the emerging photon,
$ x_-x_+ = {\hat{s}}_{\gamma\gamma}/s_{e^+e^-}$, 
with $s_{e^+e^-}$(${\hat s}_{\gamma\gamma})$ being the CM
energy squared of the $e^+e^-$($\gamma\gamma $) system, 
and $F_\pm^{\gamma/e}(P_{e^\pm}, P_{\gamma_\pm}, x_\pm; P_\pm)$
the photon distribution functions, defined in terms of
 $P_{e^\pm}, P_{\gamma_\pm}$ and $ x_\pm$ and yielding
$P_-(P_+)$, the degree of polarisation of the photon that has
back-scattered against the electron(positron)\footnote{Conventionally, 
one has $P_{-(+)}=-1(+1)$ for purely left(right)
handed photons.}.
Therefore, in terms of helicity amplitudes one has
(here, for brevity, ${\hat{\sigma}}\equiv 
{\hat{\sigma}}_{\gamma \gamma \to \sneu f \bar f^\prime}$)
\begin{eqnarray}
\nonumber
{\hat{\sigma}}(\hat s,P_-,P_+)
=&{1\over 4}&
[(1+P_-)(1+P_+){\hat{\sigma}}_{++}(\hat s)+
 (1+P_-)(1-P_+){\hat{\sigma}}_{+-}(\hat s)\\
&+&
 ~ (1-P_-)(1+P_+){\hat{\sigma}}_{-+}(\hat s) 
 + (1-P_-)(1-P_+){\hat{\sigma}}_{--}(\hat s)].
\end{eqnarray}
As polarised $\gamma$-structure functions 
we have used those of Ref. \cite{BKU}.

\end{enumerate}

The flavour of the final state fermions will depend upon the
RPV couplings involved. It has been shown that most of the first 
two generation L-violating terms are highly constrained from different
low and medium energy processes \cite{rpv_bound}. 
For our study, we made the assumption that
just one L-violating coupling at a time is the dominant one, so
that only bounds derived under the same hypothesis are relevant.
This restriction may seem  unnatural, however, it is a useful approach 
that allows one to derive 
a quantitative feeling for the phenomenological 
consequences of RPV interactions, while avoiding
a proliferation of SUSY input parameters. 
In our analysis, we will concentrate on the following L-violating
couplings: $\lambda_{311}, \lambda_{323}, \lp_{323}$ and $\lp_{333}$. The 
reason for selecting this particular set out of the 
36 possible couplings is that these  are less constrained and
at the same time can lead to a significant contribution to the production 
as well as the decay rates of 
sneutrinos in (\ref{production})--(\ref{decay}).
The upper limits on these couplings and the processes which give such bounds
are shown in Table~\ref{rpv_lim}.
\begin{table}[ht]
\begin{center}
\begin{tabular}{|c|c|c|} \hline
Coupling & Upper Limit & Sources\\
\hline
$\lambda_{311}$ & 0$.062 $ & $R_{\tau} = \frac{\Gamma(\tau \to e \nu \bar{\nu})
}{\Gamma(\tau \to e \mu \bar{\nu})} $\cite{Barger} \\
\hline
$\lambda_{323}$ & 0$.070 $  & ''\\
\hline
$\lp_{323}$ & 0$.52 $ & $ R_{D_s} = \frac{\Gamma (D_s \to \tau \nu_{\tau})}
{\Gamma (D_s \to \mu \nu_{\mu})} $\cite{greno}\\
\hline
$\lp_{333}$ & 0$.45 $ & ''\\
\hline
\end{tabular}
\end{center}
\caption{Experimental ($2\sigma $) upper bounds on the RPV couplings
         relevant to this analysis. All sfermion masses are assumed
         to be 100 GeV.}
\label{rpv_lim}
\end{table}
Notice that all these limits scale as $M_{\tilde f}/100~{\rm GeV} $ with the 
common sfermion mass. That is, they become weaker as $M_{\tilde f}$ increases. 
However, some couplings are constrained by the requirement of perturbative 
unitarity. For
example, the corresponding bound 
on $\lp_{323}$ is $1.12$. Indeed, we could have 
taken any 
values of these couplings bounded between the mentioned upper and lower 
limits (as done 
by \cite{Soni}). However, like in \cite{sourov2} and
for the sake of simplicity, we will consider only one fixed value for each of 
the RPV couplings, the one obtained assuming a $100$ GeV 
sfermion mass (as in Table~\ref{rpv_lim}). In a sense then,
our approach can be viewed as conservative. 

Once the sneutrino is produced, it will decay. Depending on its nature, 
the dominant decay modes are:
\beq
{\tilde \nu} &\to & f {\bar f^\prime}
~~~~(f = \ell, q )~~~~{\rm{fermion~pairs}},\\   
{\tilde \nu} &\to & \tilde \chi^0_i \nu ~~(i=1,2,3,4)
~~{\rm{neutralino~+~neutrino}},\\
{\tilde \nu} &\to & \tilde \chi^+_i \ell^-~~(i =1,2)
~~{\rm{chargino~+~lepton}}.
\enq  
If the sneutrino is the LSP, then it will decay through the first 
(RPV) channel,
otherwise via one of the other two (MSSM) modes. 
We show the sneutrino branching ratio (BR) into two 
fermion final states in the $\mu-M_2$ plane for a fixed value of 
$\tan\beta $, 
RPV coupling and sneutrino mass. In the course of the analysis
we assume the Grand Unification (GUT) relationship between 
the $U(1)$ and $SU(2)$ gaugino mass parameters: {\it i.e.},
\beq
M_1 = \frac{5}{3}\tan^2\theta_W M_2.
\enq  
Hence, the sneutrino BR into two fermions will depend 
upon $\mu, M_2, \tan\beta$, $M_{\tilde \nu}$ and the magnitude of the 
RPV coupling.  To study the variation of the sneutrino RPV BR
we have spanned $\mu$ from $-500$ GeV to $+500$ GeV and $M_2$ 
from $100$ GeV to 500 GeV.

\begin{figure}[hbt]
\centerline{\epsfig{file=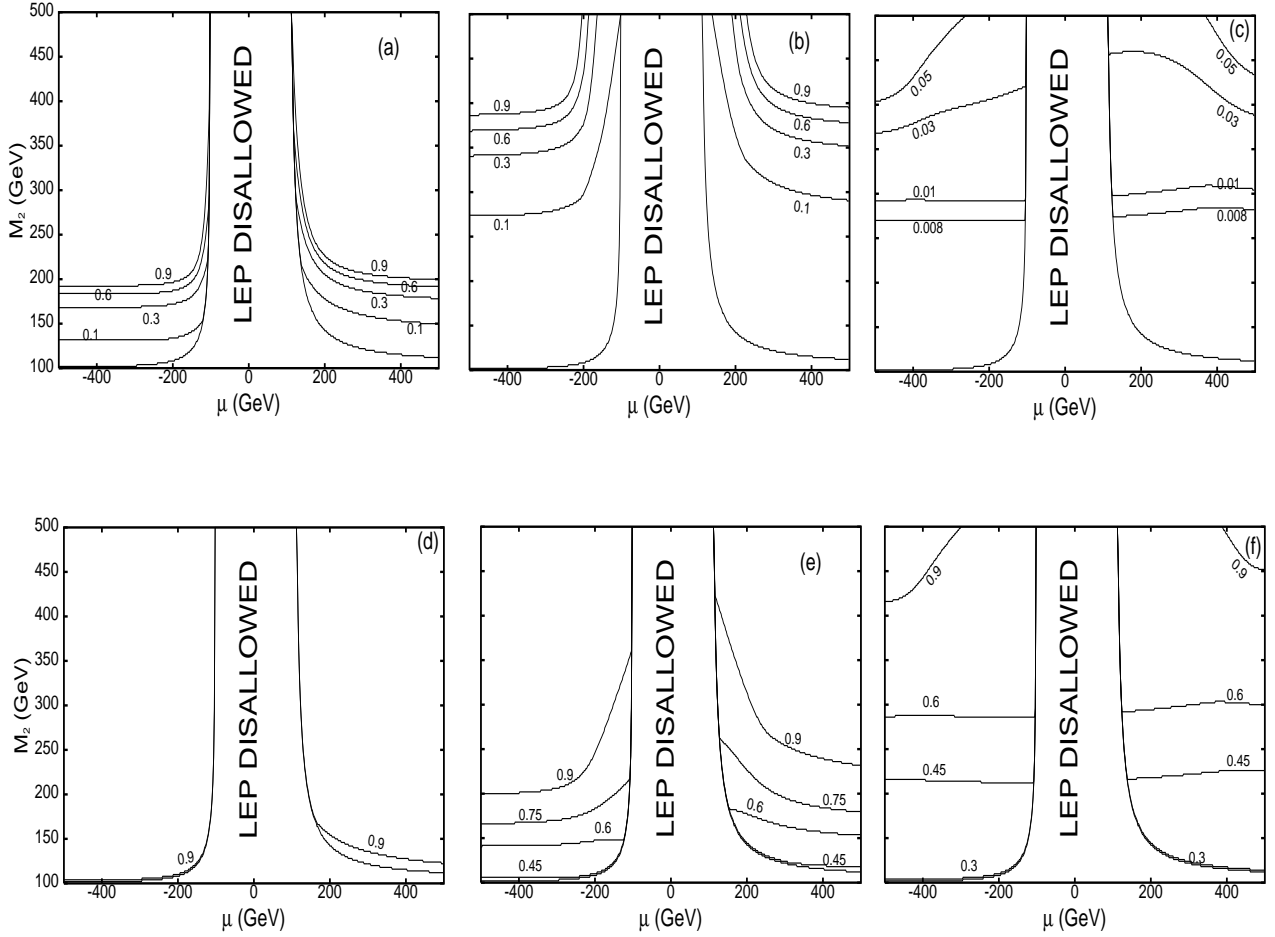,width=\linewidth}}
\vspace*{-6.0cm}
\caption{Constant BR
         contours of the decay $\tilde \nu_{\tau} \to e^+ e^-$ for three
         values of sneutrino masses: 100 GeV $(a)$, 200 GeV $(b)$ and
         400 GeV $(c)$. Figures (d)--(f) represents contours of constant
         BR($\tilde \nu_{\tau} \to  b \bar b $), again for a
         100, 200 and 400 GeV sneutrino mass, respectively.
         The relevant L-violating couplings are here:
         $\lambda_{311} = 0.062$ for (a)--(c)
         and $\lambda^\prime_{333} = 0.45$ for (d)--(f). 
         The level curves are function of $\mu$ and $M_2$, whereas
         the other relevant MSSM parameters are those of set A in
         Table.~\ref{set_mssm} below.}
\label{br_cont}
\end{figure}

In Figure~\ref{br_cont}(a) we show the contours of constant BR(${\tilde \nu_{\tau}} 
\to e^+ e^-$) through the $\lambda_{311}$ coupling
for $ M_{\tilde \nu_{\tau}}=100$ GeV  in the $\mu-M_2$ plane, with
$\tan\beta = 5$. The region labelled by `{LEP DISALLOWED}' is ruled out from
the kinematic limit on the lighter chargino mass extracted
from LEP-2 data. It can be
seen from this Figure that the mentioned BR is $90\%$ over a large 
part of the parameter space. In this case, the lighter chargino is heavier 
than the sneutrino mass, forbidding the 
${\tilde \nu} \to \tilde \chi^+_1 \ell^- $ decay channel. The only
MSSM channel allowed is ${\tilde \nu} \to  \tilde \chi^0_1 \nu$, which
dominates in the low $M_2$ region, where $M_{\tilde \chi^0_1}< M_{\tilde\nu}$.
The above scenario changes once the sneutrino becomes heavier,
 as shown in Figure~\ref{br_cont}(b), where the same BR as above is plotted
but now with $M_{\tilde\nu}=200$ GeV. In this case, both channels
${\tilde \nu} \to  \tilde \chi^+_1 \ell^- $ and
${\tilde \nu} \to  \tilde \chi^0_1 \nu$ make a
significant contribution
to the total decay width of the sneutrino. (The RPV BR
increases with $M_2$ though,
since the lighter chargino and neutralino become heavier.) 
In Figure~\ref{br_cont}(c), this trend
becomes very clear: for a 400 GeV sneutrino most
of the $\mu -M_2$ plane is covered by the MSSM decays, relegating large
RPV BRs to small corners of the  parameter space.

This situation changes considerably when the
RPV coupling is $\lp_{333}$. In this case, because of the 
larger magnitude of the latter, as compared to $\lambda_{311}$, the 
BR$({\tilde \nu} \to b \bar b)$ for a 100 GeV sneutrino mass covers 
almost the entire $\mu - M_2$ plane analysed in this paper. Even 
for heavier
sneutrinos ({\it e.g.}, 200 GeV and 400 GeV), a larger area in the
$\mu - M_2$ plane is dominated
by the above BR, leaving a smaller region for the MSSM decays than
in the previous case: see Figures~\ref{br_cont}(d)--(f). Finally, we have
noticed that this general behavior of the BRs does not change for 
higher values of $\tan\beta$. Also,
the impact of $\lambda_{323}$ and $\lambda'_{323}$ 
RPV couplings onto the decay rates induces a pattern similar to the
one discussed, so we do not reproduce the corresponding Figures here.

\section{Numerical analysis}

We perform our numerical analysis for three different points in the
MSSM parameter space allowed by LEP-2 data.
These are representative of three different
natures of the lightest chargino and are
defined in Table~\ref{set_mssm}. 

\begin{table}[ht]
\begin{center}
\begin{tabular}{|c|c|c|c|c|c|c|} \hline
Set & $\mu $~(GeV) & $M_2$~(GeV)& $\tan\beta $ & $M_{\tilde\chi^0_1}$~(GeV) & 
$M_{ \tilde\chi^\pm_1} $~(GeV) & Nature of $\tilde\chi^\pm_1 $\\
\hline
{A} & $-400$ & 150 & 5 & 76. 4 & 150.3 & Gaugino dominated state \\
\hline
{B} & $~200$ & 350 & 40 & 150.4 & 185.6 & Mixed state \\
\hline
{C} & $~175$ & 500 & 40 & 155.6 & 169.4 & Higgsino dominated state \\
\hline
\end{tabular}
\end{center}
\caption{Set of selected points in the MSSM parameter space with LSP 
and lighter chargino mass (and nature) given explicitly (we 
defer to the Appendix the listing 
of the total decay widths of the sneutrino in different RPV channels for 
these three choices of MSSM parameters). }
\label{set_mssm}
\end{table}

Furthermore, we select the combinations of incident laser
and electron beam polarisations shown in Table~\ref{beam_pol}.

\begin{table}[ht]
\begin{center}
\begin{tabular}{|c|c|c|c|c|} \hline
$\;$  & $P_{\gamma_+}$ & $P_{\gamma_-}$ & $P_{e^+}$ & $P_{e^-}$ \\
\hline
$\sigma(+-)$ & $+1$ & $-1$ & $-0.8$ & $+0.9$ \\
\hline
$\sigma(++)$ & $+1$ & $+1$ & $-0.8$ & $-0.9$ \\
\hline
$\sigma(00)$ & $0$ & $0$ & $0$ & $0$ \\
\hline
\end{tabular}
\end{center}
\caption{Values of laser and electron(positron) beam polarisations
adopted in our analysis. The $\sigma(+-)$
and $\sigma(++)$ denote the corresponding polarised
production cross-sections, 
with $\sigma(00)$ the unpolarised one.} 
\label{beam_pol}
\end{table}

The choice $P_{\gamma_\pm}P_{e^\pm}< 0 $ 
guarantees not only good mono-chromaticity, but 
also a high degree of circular polarisation of the produced photons
as compared to the case $P_{\gamma_\pm}P_{e^\pm}> 0 $. There exists a
symmetry amongst the four 
combinations of laser polarizations, as
 $(+-)$ and $(-+)$ give the same 
result, and so do 
$(++)$ and $(--)$ (see also \cite{BKU}).  

To mimic the finite coverage of the LC detectors, we impose the following
cuts on the final state particles in (\ref{production})\footnote{We identify
jets with the partons from which they originate.}:
\beq\label{cuts}
5^{\rm o} & <&  \theta ~~< ~~175^{\rm o}~~~
({\rm angular~cut~on~both~leptons~and~jets }),\\
E_{\ell} & >&  5~{\rm GeV}~~~({\rm energy~ cut~ on~ leptons}),\\
E_{j}    & >&  10~{\rm GeV}~~~({\rm energy~ cut~ on~ jets}).
\enq



As already mentioned, we assume that only one between the $\lambda$
and $\lambda'$ couplings dominates at a time. Besides, we will treat the
signatures arising from the four RPV couplings considered here, {\it i.e.},
$\lambda_{311}, \lambda_{323}, \lp_{323}$ and $\lp_{333}$, separately
in the four subsections below. Where appropriate, all possible
electromagnetic (EM)
charge combinations (c.c.'s) will be included. Moreover, we assume
that the EM charge of the leptons 
($e,\mu$ and $\tau$) can always be determined, unlike the case of 
quarks. For the latter, we will assume a benchmark 100\% efficiency
in tagging $b$ flavours.

\subsection{Signals from the $\lambda_{311}$ coupling} 

Presence of this coupling leads $\sneu_{\tau}$ to decay into $e^+e^-$
pairs. Hence, the signal corresponding to this L-violating coupling is
$e^+e^- e^+e^-$. In Figure~\ref{signal_1}(a) we show the variation of
$\sigma (\gamma\gamma \to \sneu_{\tau} e^+e^-)
*{\rm{BR}}(\sneu_{\tau}\to e^+e^- )$ as a function of the
$ \sneu_{\tau}$ mass for the MSSM set {A}, at $\sqrt{s_{e^+e^-}}= 500$~GeV.
The effect of beam polarisation can be seen very clearly from the figure. At 
very low sneutrino masses ($< 150$--200 GeV), $\sigma(++)$, $\sigma(+-)$ 
and the unpolarised cross-section $\sigma(00)$
are basically the same. As the sneutrino mass rises, the above three
cross-section display a hierarchy, though not dramatic, 
with $\sigma(+-) > \sigma(00) > \sigma(++)$, whereas, for  
$M_{\sneu_{\tau}}\geq 0.5\sqrt{s_{e^+e^-}}$, the  $\sigma(++)$ component 
is the one which largely dominates. A 
similar situation can be seen for the other two sets of MSSM parameters,
namely sets {B} and {C}, in 
Figures~\ref{signal_2}(a) and \ref{signal_3}(a), respectively. 
For $\sqrt{s_{e^+e^-}} = 1$ TeV, corresponding plots are given
in Figures~\ref{signal_4}(a), 
\ref{signal_5}(a) and \ref{signal_6}(a), for the MSSM
parameter sets {A}, {B} and {C}, respectively. At higher energies, the
pattern is very similar, with the only exceptions that in
this case  $\sigma(00)$ is slightly larger than the other two at small
sneutrino masses and the mentioned hierarchy onsets for somewhat larger
values of the latter, in comparison to the lower energy collider option. 

\subsection{Signals from the $\lambda_{323}$ coupling} 

\begin{figure}[hbt]
\centerline{\epsfig{file=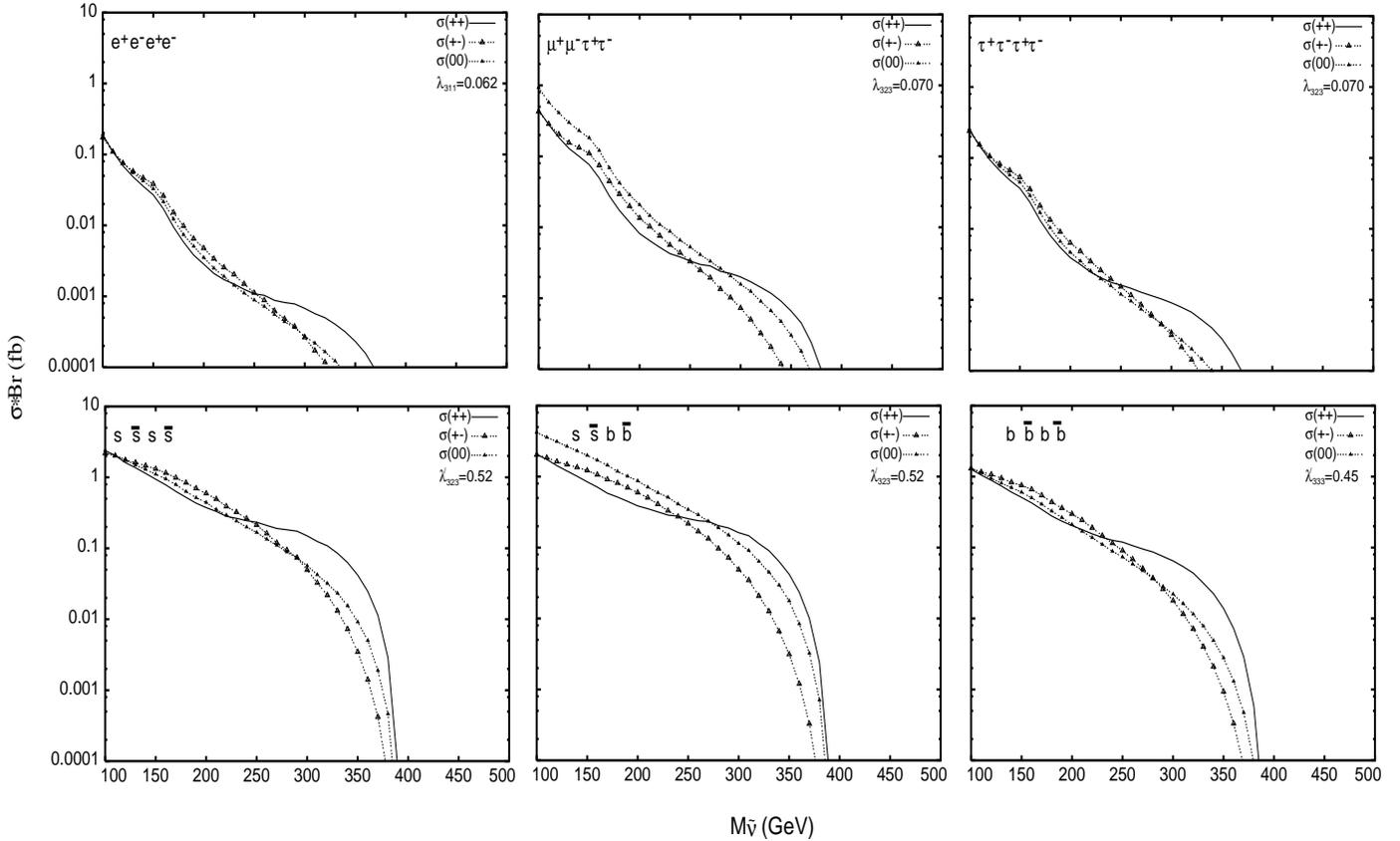,width=\linewidth}}
 \vspace*{-7.5cm}
\caption{Variation of $\sigma(\gamma \gamma \to \tilde \nu_i f_j \bar f_k) 
*{\rm{BR}}(\tilde \nu_{i} \to f_j \bar f_k)$ at 
$\sqrt{s_{e^+e^-}}= 500$~GeV with the
sneutrino mass, for fixed values of the relevant
$\lambda_{ijk}$ and $\lp_{ijk}$ couplings. 
The MSSM parameters are $\mu = -400~{\rm GeV}$, 
$M_2= 150~{\rm GeV}$ and $\tan\beta=5$ (set A). 
Final state fermions are shown in the respective figures. For dissimilar 
fermions in the final state we include all charge conjugate states.
See Table 3 for the definition of (un)polarised cross-sections.} 
\label{signal_1}
\end{figure}

\begin{figure}[hbt]
\centerline{\epsfig{file=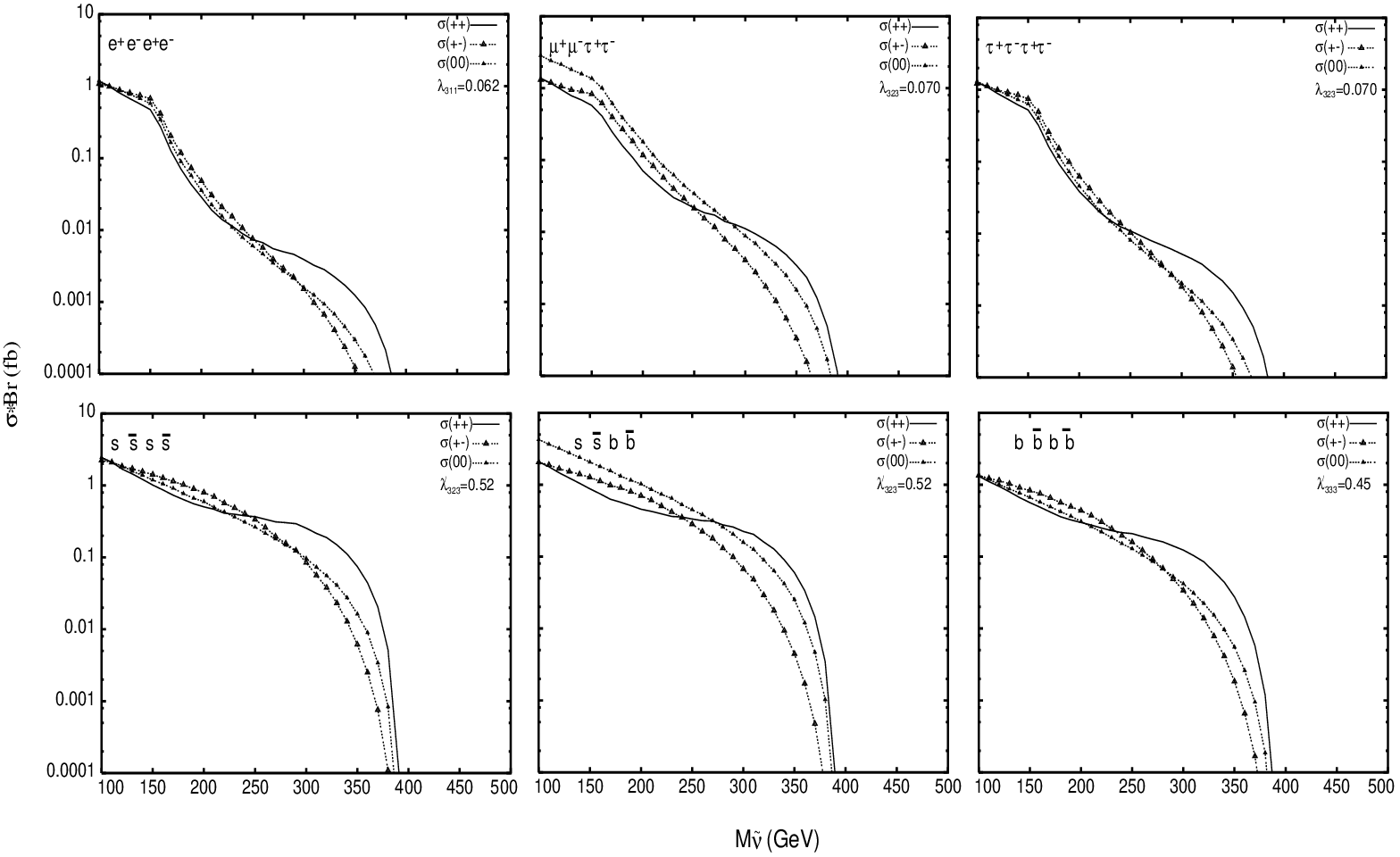,width=\linewidth}}
 \vspace*{-7.5cm}
\caption{Variation of $\sigma(\gamma \gamma \to \tilde \nu_i f_j \bar f_k) 
*{\rm{BR}}(\tilde \nu_{i} \to f_j \bar f_k)$ at $\sqrt{s_{e^+e^-}}= 500$~GeV 
with the sneutrino mass, for fixed values of the relevant
$\lambda_{ijk}$ and $\lp_{ijk}$ couplings. The MSSM parameters are 
$\mu = 200~{\rm GeV}$, $M_2= 350~{\rm GeV}$ and
$\tan\beta=40$  (set B). 
Final state flavours are shown in the each figure. We inlcude the C.C. states
for the dissimilar final states.}
\label{signal_2}
\end{figure}

\begin{figure}[hbt]
\centerline{\epsfig{file=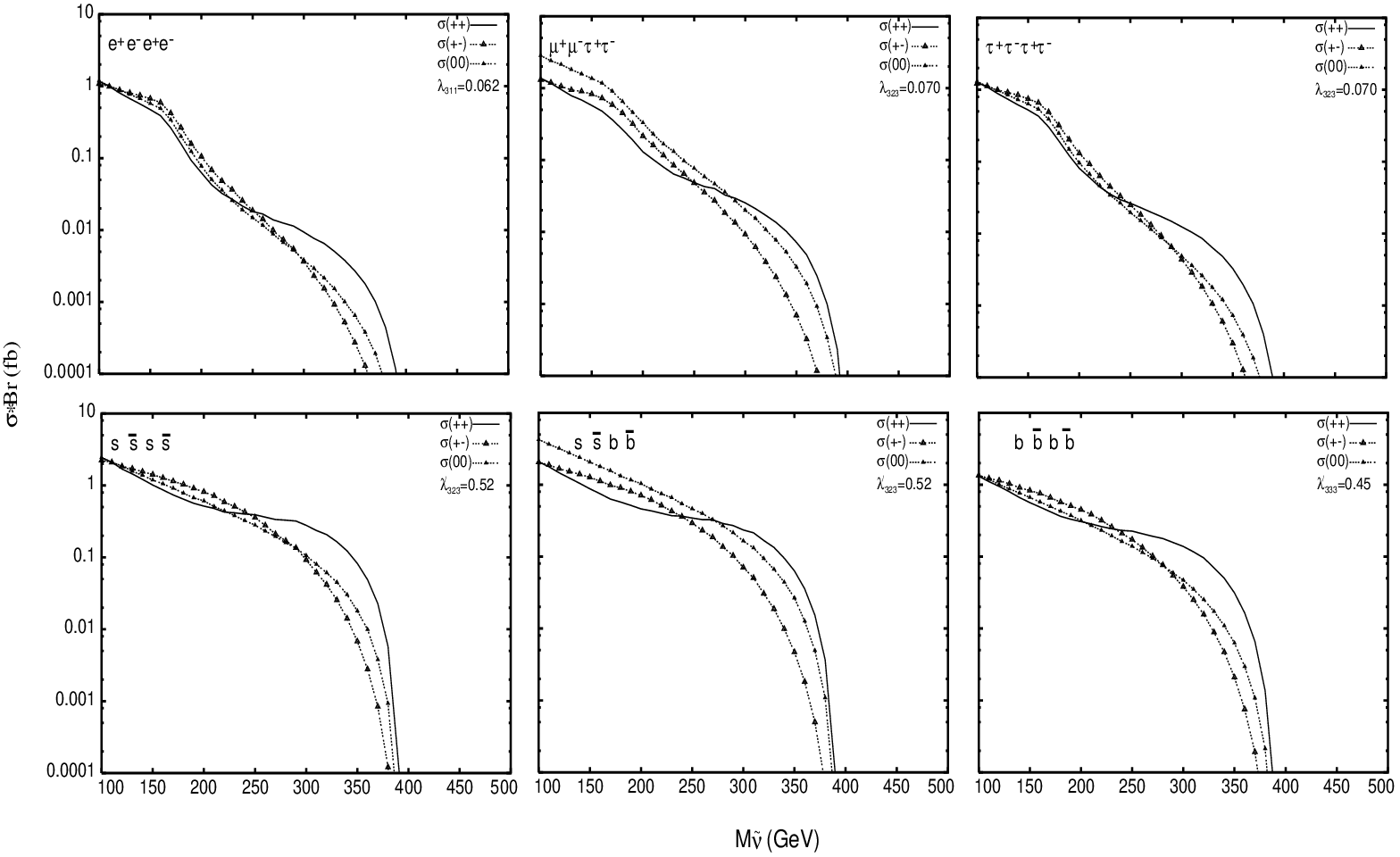,width=\linewidth}}
 \vspace*{-7.5cm}
\caption{Variation of $\sigma(\gamma \gamma \to \tilde \nu_i f_j \bar f_k) 
*{\rm{BR}}(\tilde \nu_{i} \to f_j \bar f_k)$ at $\sqrt{s_{e^+e^-}}= 500$~GeV 
with the
sneutrino mass, for fixed values of the relevant
$\lambda_{ijk}$ and $\lp_{ijk}$ couplings. The MSSM parameters are 
$\mu = 175~{\rm GeV}$, 
$M_2= 500~{\rm GeV}$ and
$ \tan\beta=40$  (set C). 
Final state flavours are shown in the each figure. We inlcude the C.C. states
for the dissimilar final states.}
\label{signal_3}
\end{figure}

\begin{figure}[hbt]
\centerline{\epsfig{file=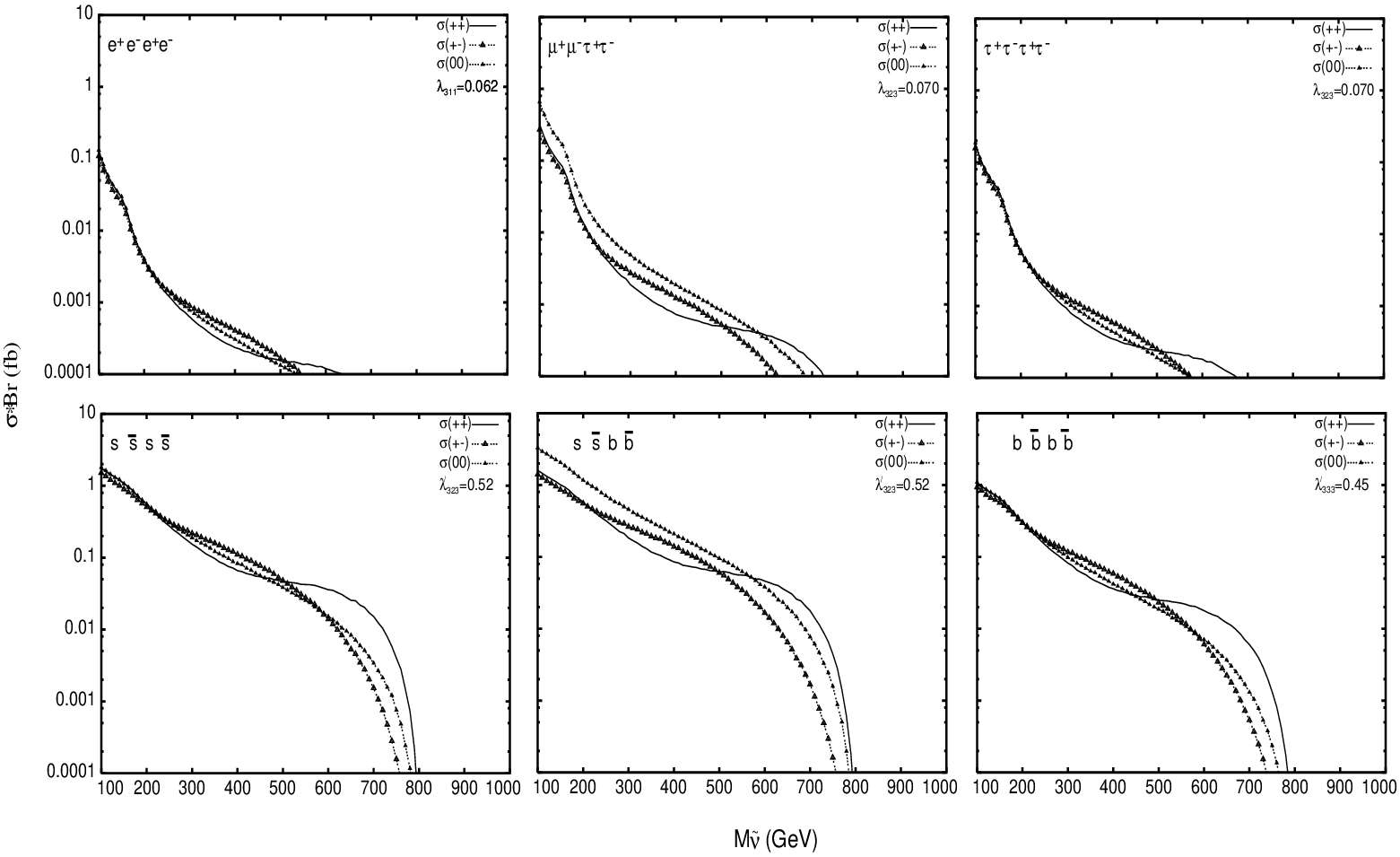,width=\linewidth}}
 \vspace*{-7.5cm}
\caption{Variation of $\sigma(\gamma \gamma \to \tilde \nu_i f_j \bar f_k) 
*{\rm{BR}}(\tilde \nu_{i} \to f_j \bar f_k)$ at $\sqrt{s_{e^+e^-}}= 1$~TeV 
with the
sneutrino mass, for fixed values of the relevant
$\lambda_{ijk}$ and $\lp_{ijk}$ couplings. 
The MSSM parameters are $\mu = -400~{\rm GeV}$, 
$M_2= 150~{\rm GeV}$ and
 $\tan\beta=5$ (set A). 
Final state flavours are shown in the each figure. We inlcude the C.C. states
for the dissimilar final states.}
\label{signal_4}
\end{figure}

\begin{figure}[hbt]
\centerline{\epsfig{file=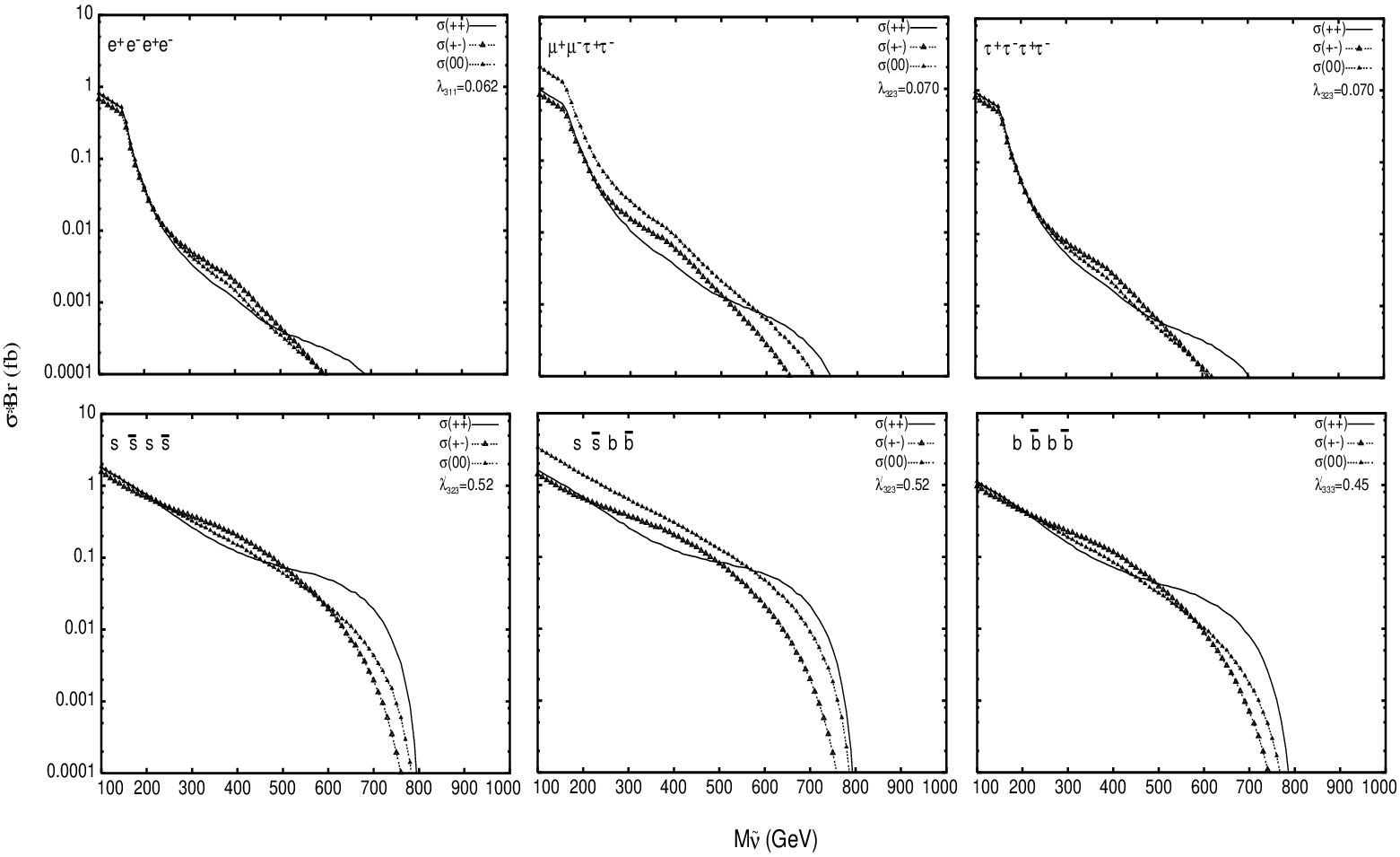,width=\linewidth}}
 \vspace*{-7.5cm}
\caption{Variation of $\sigma(\gamma \gamma \to \tilde \nu_i f_j \bar f_k) 
*{\rm{BR}}(\tilde \nu_{i} \to f_j \bar f_k)$ at $\sqrt{s_{e^+e^-}}= 1$~TeV 
with the
sneutrino mass, for fixed values of the relevant
$\lambda_{ijk}$ and $\lp_{ijk}$ couplings. 
The MSSM parameters are $\mu = 200~{\rm GeV}$, 
$M_2= 350~{\rm GeV}$ and $\tan\beta=40$  (set B). 
Final state flavours are shown in the each figure. We inlcude the C.C. states
for the dissimilar final states.}
\label{signal_5}
\end{figure}

\begin{figure}[hbt]
\centerline{\epsfig{file=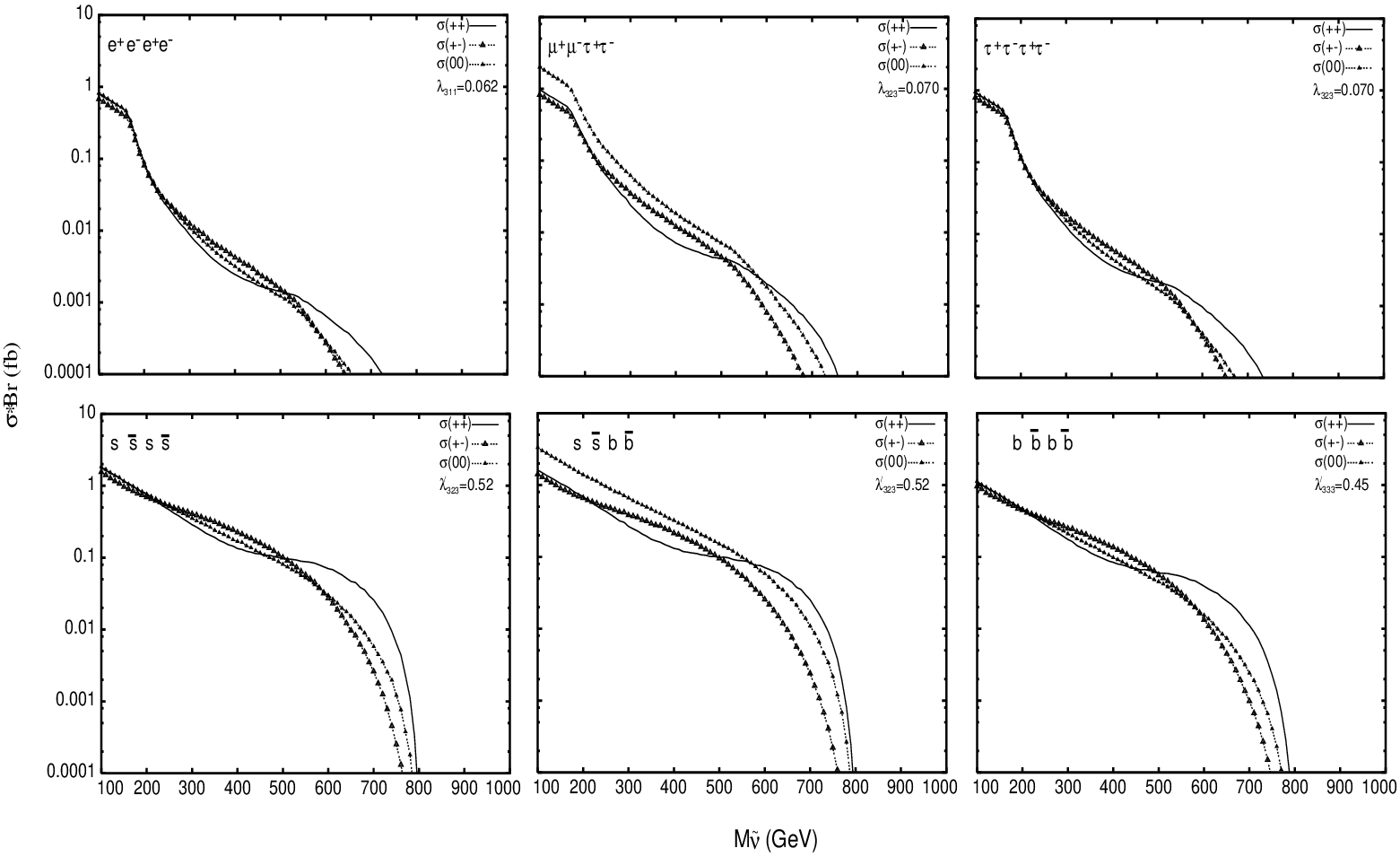,width=\linewidth}}
 \vspace*{-7.5cm}
\caption{Variation of $\sigma(\gamma \gamma \to \tilde \nu_i f_j \bar f_k) 
*{\rm{BR}}(\tilde \nu_{i} \to f_j \bar f_k)$ at $\sqrt{s_{e^+e^-}}= 1$~TeV with
the sneutrino mass, for fixed values of the relevant
$\lambda_{ijk}$ and $\lp_{ijk}$ couplings. 
The MSSM parameters are $\mu = 175~{\rm GeV}$, 
$M_2= 500~{\rm GeV}$ and $ \tan\beta=40$ (set C). 
Final state flavours are shown in the each figure. We inlcude the C.C. states
for the dissimilar final states.}
\label{signal_6}
\end{figure}

Presence of this coupling gives rise to the following two 
types of signals: flavour conserving $\tau^+\tau^-\tau^+\tau^-$ and  
flavour changing $ \mu^+\mu^- \tau^+\tau^- $ (and c.c.'s). 
The variation of 
$\sigma (\gamma\gamma \to \sneu_{\tau}  \mu^+\tau^-)
*{\rm{BR}}(\sneu_{\tau}\to \mu^-\tau^+)$ as a function of the
sneutrino mass is shown in Figures~\ref{signal_1}(b)
\ref{signal_2}(b) and \ref{signal_3}(b),
 for $\sqrt{s_{e^+e^-}}=500$ GeV, and 
Figures~\ref{signal_4}(b), \ref{signal_5}(b) and \ref{signal_6}(b) 
for $\sqrt{s_{e^+e^-}} = 1$ TeV, corresponding to the MSSM
parameter sets {A}, {B} and {C}, respectively. 
In this case the final state will have  
three different combinations of charged particles
with identical rates: $ \mu^+\mu^- \tau^+\tau^- $,
$\mu^+\mu^+\tau^-\tau^-$ and $\mu^-\mu^-\tau^+\tau^+$.  Hence, the
individual channels will be $1/3$ of the total cross-section shown in the
Figures. 
%
The plots for the flavour conserving final states are displayed in Figures~
\ref{signal_1}(c),
\ref{signal_2}(c) and \ref{signal_3}(c), for $\sqrt{s_{e^+e^-}}=500$ GeV, 
and 
Figures~\ref{signal_4}(c), \ref{signal_5}(c) and \ref{signal_6}(c), 
for $\sqrt{s_{e^+e^-}} = 1$ TeV.

In this case too we see that the dominant cross-section comes from 
$\sigma(++)$ once the $M_{\sneu_{\tau}} \geq 0.5 \sqrt{s_{e^+e^-}} $.
However, at lower sneutrino masses, the pattern is different from the
previous case. The ordering 
$\sigma(+-) > \sigma(00) > \sigma(++)$
in the intermediate mass regime and the convergence
of the rates for all polarisation states at small $M_{\sneu_{\tau}}$
values hold only for $\tau^+\tau^-\tau^+\tau^-$, not for
$\mu^+\mu^-\tau^+\tau^-$ (plus c.c.s), 
for which the unpolarised cross sections are always largest. 
In this case, again, the increase in CM energy delays the onset of
the highlighted cross section hierarchy, for $\tau^+\tau^-\tau^+\tau^-$
final states.

\subsection{Signals from the $\lp_{323}$ coupling} 

Presence of this coupling gives rise to the following three 
types of signals: the flavour conserving $s \bar s s \bar s $
and $b \bar b b \bar b $ plus the 
flavour changing $ s {\bar s} b {\bar b}  $ (and c.c.'s). The variation of 
$\sigma (\gamma\gamma \to \sneu_{\tau}  b {\bar s} )
*{\rm{BR}}(\sneu_{\tau}\to  {\bar b} s )$ as a function of the sneutrino 
mass is shown in Figures~\ref{signal_1}(e),
\ref{signal_2}(e) and \ref{signal_3}(e), for $\sqrt{s_{e^+e^-}}=500$ 
GeV, and 
Figures~\ref{signal_4}(e), \ref{signal_5}(e) and \ref{signal_6}(e) 
for $\sqrt{s_{e^+e^-}} = 1$ TeV, again, in correspondence of the MSSM
parameter sets {A}, {B} and {C}, respectively. 
Notice that in this case too there are three equiprobable signatures:
$ s s {\bar b} {\bar b} $,
$s {\bar s} b {\bar b} $ and ${\bar s} {\bar s}b b $. Corresponding
plots for the flavour conserving modes are displayed in Figures
\ref{signal_1}(d),
\ref{signal_2}(d) and \ref{signal_3}(d), for $\sqrt{s_{e^+e^-}}=500$ GeV, 
and 
Figures~\ref{signal_4}(d), \ref{signal_5}(d) and \ref{signal_6}(d) 
for $\sqrt{s_{e^+e^-}} = 1$ TeV (in correspondence of sets A,B and C).

The dependence upon the beam polarisation configuration is basically the same
as the one described in the previous section, once one establishes a
correspondence between the identical- and different-flavour final states
in the two cases. The energy dependence does not differ much  either from
that in the two previous cases. 

\subsection{Signals from the $\lp_{333}$ coupling} 

Presence of this coupling will also give rise to the 
signal $b \bar b b \bar b $. The numerical results for the
corresponding production cross-sections are shown in  
Figures~\ref{signal_1}(f),
\ref{signal_2}(f) and \ref{signal_3}(f), for $\sqrt{s_{e^+e^-}}=500$ GeV, 
and Figures~\ref{signal_4}(f), \ref{signal_5}(f) and \ref{signal_6}(f) 
for $\sqrt{s_{e^+e^-}} = 1$ TeV, corresponding to the MSSM
parameter sets {A}, {B} and {C}, respectively. 

As for the beam polarisation dependence, here, one can see the
usual dominance of $\sigma(++)$ whenever   
$M_{\sneu_{\tau}}\geq 0.5\sqrt{s_{e^+e^-}}$, 
with the  $\sigma(+-)$ component dominating in the
intermediate regime. For lower masses, the energy dependence is
such that at 500 GeV $\sigma(+-)$ is above $\sigma(00)$,
whereas at 1 TeV things go the other way around.

\subsection{Signals from $ \sneu \to \tilde \chi^+_1 \ell^- $ } 

\begin{figure}[hbt]
\centerline{\epsfig{file=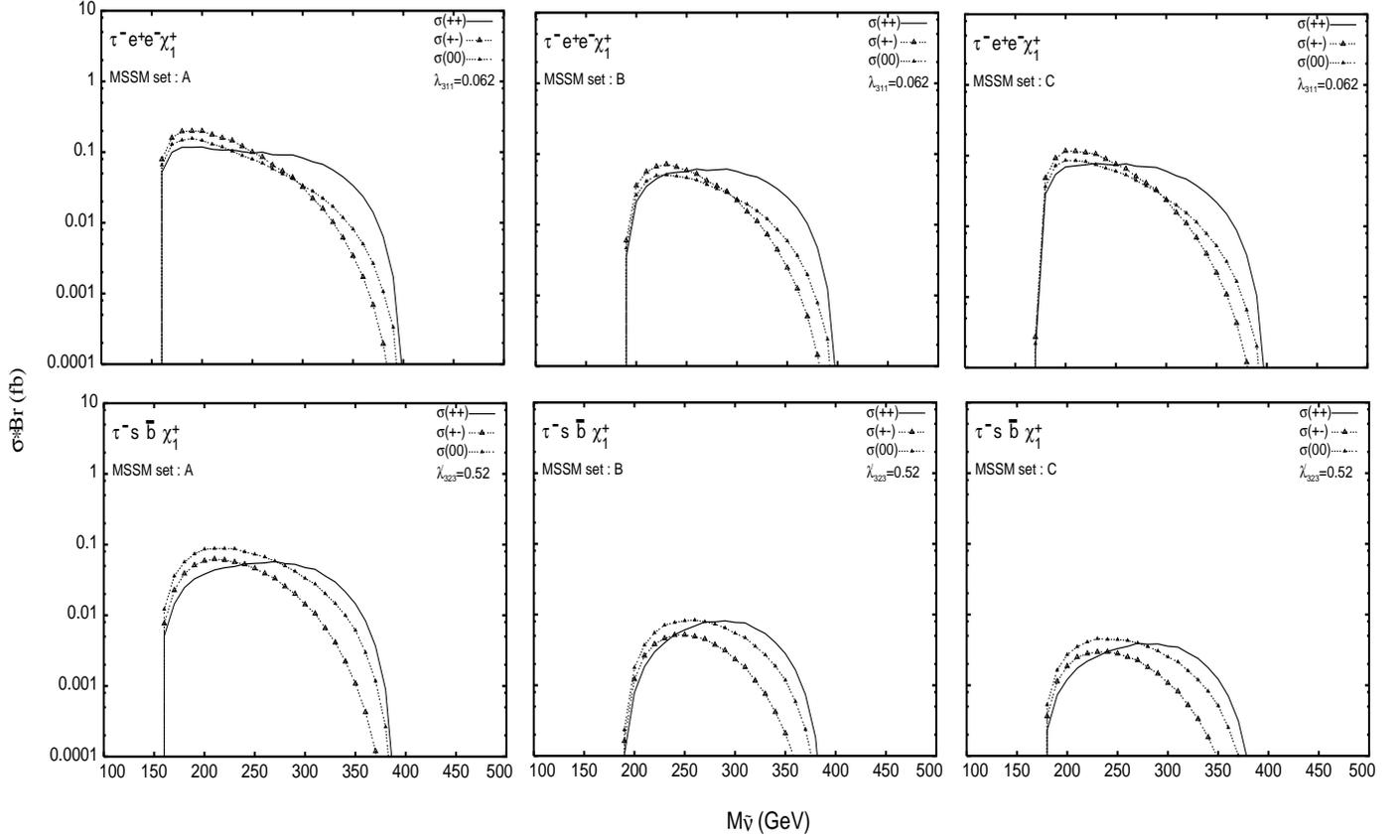,width=\linewidth}}
 \vspace*{-7.5cm}
\caption{Variation of $\sigma(\gamma \gamma \to \tilde \nu_i f_j \bar f_k) 
*{\rm{BR}}(\tilde \nu_{i} \to \ell^- \tilde \chi^+_1)$ at 
$\sqrt{s_{e^+e^-}}= 500 $~GeV with the
sneutrino mass, for fixed values of the relevant
$\lambda_{ijk}$ and $\lp_{ijk}$ couplings. 
MSSM parameter sets are shown in each figurs. Beam 
polarization conventions are the same as in the previous Figures.}
\label{signal_7}
\end{figure}

Here, we would like to comment about the signal cross-section 
$\sigma (\gamma \gamma \to \sneu f \bar f ) * 
{\rm{BR}}(\sneu \to \tilde \chi^+_1 \ell^-) $ for two different 
RPV interactions, namely $\lambda_{311}$ and $\lp_{323}$.  
Figures~\ref{signal_7}(a)--(c)
correspond to $\sigma (\gamma \gamma \to \sneu_{\tau} e^+ e^-)
*{\rm{BR}}(\sneu_{\tau} \to \tilde \chi^+_1 \tau^-) $ 
for
$\lambda_{311}=0.062$ whereas the
variation of  $\sigma (\gamma \gamma \to \sneu s{\bar b}) 
*{\rm{BR}}(\sneu_{\tau} \to \tilde \chi^+_1 \tau^-) $ with the 
sneutrino mass (for $\lp_{323}= 0.52$)
is shown in  Figures~\ref{signal_7}(d)--(f). Notice that
Figures~\ref{signal_7}(a,d), \ref{signal_7}(b,e) and 
\ref{signal_7}(c,f) correspond to the three usual sets of
MSSM parameters {A}, {B} and {C}, respectively. These 
cross-sections have been calculated for the case of a LC of 500 GeV.
The pattern of the production and decay rates is 
here quite different from the one displayed
for the case of RPV decays of the sneutrino. In fact,  
the overall behaviour in this channel depends on other
factors. Firstly, on the relative mass difference between $\sneu$ and 
$\tilde\chi^+_1$, as well as upon 
the composition of $\chi^+_1$ (if it is Higgsino dominated, 
then the $\sneu-\tilde\chi^+_1-\ell^-$ coupling will be Yukawa suppressed).
Secondly, and most importantly, the magnitude of the
RPV coupling involved: as it is clear from comparing   
Figures~\ref{signal_7}(a)--(c) to Figure~\ref{signal_7}(d)--(f), the stronger 
the RPV coupling the smaller the 
$\tilde\nu \rightarrow \tilde\chi^+_1 \ell^-$ decay mode. In other words, this
signal is somehow complementary to the RPV ones discussed so far 
and requires a different discussion of the decay dynamics, given the
additional dependence on the chargino mass. Hence, although this signature
may well induce visible events in the end, we do not pursue 
further its study here. 

\subsection{The SM irreducible background}

If $M_{\tilde \nu}$ is very near the EW scale, say, 80--90 GeV, it
is clear that the dominant SM irreducible background to RPV
signals of the type discussed in the previous sections arises from
associated production of a $Z$ boson and a fermionic pair, with
the gauge boson decaying into two further fermions:
\begin{equation}\label{Z}
\gamma \gamma \to Z \ell^\pm \ell^{\mp}~~{\rm{or}}~~ 
{Z} q {\bar q }
\end{equation}
{\rm{with}}
\begin{equation}\label{Zdecay}
{Z} \to \ell^{'\pm} \ell^{'\mp}~~{\rm{or}}~~ 
{Z} \to q' {\bar q}^{'}.
\end{equation}
Only in the case 
of four-quark final states one has to deal with $W^\pm$ mediated
production:
\begin{equation}\label{W}
\gamma \gamma \to W^\pm q {\bar q' }
\end{equation}
{\rm{with}}
\begin{equation}\label{Wdecay}
{W^\pm} \to q'' {\bar q}^{'''}. 
\end{equation}

\begin{table}[ht]
\begin{center}
\begin{tabular}{|c|c|c|c|c|c|c|} \hline
$~$&
$e^+e^- e^+e^-$ &
$\mu^+ \mu^- \tau^+\tau^{- *}$ &
$\tau^+ \tau^- \tau^+ \tau^-$ &
$s \bar s s \bar s$ &
$s \bar s b\bar b^*$ &
$b \bar b  b \bar b$ \\ \hline\hline
$\sigma(+-)$ & $4.7\times10^{9}$ & 2169 & 232 & 532048 & 3409 & 313 \\ \hline
$\sigma(++)$ & $4.7\times10^{9}$ & 2288 & 233 & 551620 & 3485 & 317 \\ \hline
$\sigma(00)$ & $4.7\times10^{9}$ & 2228 & 232 & 541824 & 3447 & 315 \\ \hline
\multicolumn{7}{|c|}
{$\sqrt{s_{e^+e^-}}=500$ GeV}\\ \hline\hline
$\sigma(+-)$ & $1.9\times10^{9}$ & 1054 & 114 & 241093 & 1629 & 161 \\ \hline
$\sigma(++)$ & $1.9\times10^{9}$ & 1042 & 115 & 240166 & 1707 & 162 \\ \hline
$\sigma(00)$ & $1.9\times10^{9}$ & 1048 & 114 & 240612 & 1668 & 161 \\ \hline
\multicolumn{7}{|c|}
{$\sqrt{s_{e^+e^-}}=1$ TeV}\\ \hline\hline
\multicolumn{7}{|c|}
{$^*$Other c.c.'s are free from SM background}\\ \hline
\end{tabular}
\end{center}
\caption{Cross sections in femtobarns 
for the full four-fermion SM processes discussed in the text, 
for the three beam
polarisation configurations in Table~\ref{beam_pol}, after the
cuts in (\ref{cuts}) (and $M_{e^+e^-}>1$ GeV for the $e^+e^-e^+e^-$
final state). Notice that no summation
over $u,d$ and $s$ (light) flavours has been performed in the case of 
signatures involving $s$ quarks.}
\label{backgrounds}
\end{table}

\begin{table}[!t]
\begin{center}
\begin{tabular}{|c|c|c|c|c|c|c|} \hline
$M_{\sneu}$ (GeV) &
$e^+e^- e^+e^-$ &
$\mu^+ \mu^- \tau^+\tau^{- *}$ &
$\tau^+ \tau^- \tau^+ \tau^-$ &
$s \bar s s \bar s$ &
$s \bar s b\bar b^*$ &
$b \bar b  b \bar b$ \\ \hline\hline
$100$ & 0.1908 & 0.4650 & 0.2565 & 2.387  & 2.100  & 1.3091 \\ 
      & 1.1865 & 1.4177 & 1.3095 & 2.4466 & 2.1270 & 1.3529       \\ 
      & 1.1865 & 1.4177 & 1.3095 & 2.4466 & 2.1270 & 1.3529       \\ 
      & 382042836 & 56  & 19  & 93366 & 627 & 63  \\ \hline
$200$ & 0.00234 & 0.00695 & 0.00313  & 0.3870 & 0.4019  & 0.17967 \\ 
      & 0.03256 & 0.08342 & 0.04208  & 0.5939 & 0.51151 & 0.30668         \\ 
      & 0.6004  & 0.13705 & 0.07569  & 0.6055 & 0.51651 & 0.31462        \\ 
      & 93921298  & 6.4 & 4.4 & 22248 & 170 & 18  \\ \hline
$300$ & 0.000171& 0.000493 & 0.000222 & 0.04387  & 0.0477   & 0.01649\\ 
      & 0.00112 & 0.003163 & 0.001461 & 0.09135  & 0.07628  & 0.03892   \\ 
      & 0.00267 & 0.007204 & 0.003431 & 0.10293  & 0.08134  & 0.04533    \\ 
      & 19052146  & 3.6 & 1.7 & 3696  & 30  & 3.3 \\ \hline
\multicolumn{7}{|c|}
{$^*$Other c.c.'s are free from SM background}\\ \hline
\end{tabular}
\end{center}
\caption{Cross sections in femtobarns 
for the four-fermion final states discussed in the text,
for both signals (first three rows for MSSM set A,B and C respectively) 
and backgrounds (last row),  
for unpolarised beams at ${\sqrt{s_{e^+e^-}}} = 500 $~GeV , after the
cuts in (\ref{cuts}) plus the additional constraint $|M_{ff}-M_{\sneu}|<5$
GeV, for several sneutrino masses. (Here, $ff$ refer to either
lepton-lepton or jet-jet pairs and only one  $M_{\sneu}$ is required
to be reconstructed.) Notice that, for the backgrounds, no summation
over $u,d$ and $s$ (light) flavours has been performed in the case of 
signatures involving $s$ quarks.}
\label{mass_cuts}
\end{table}
However, notice that, 
with the exception of the $s\bar s s\bar s$ signature, only 
Cabibbo-Kobayashi-Maskawa (CKM) suppressed channels can contribute 
in (\ref{W})--(\ref{Wdecay}), if one assumes a fully efficient 
heavy quark tagging (via a displaced vertex)
to be available at future LCs ({\it i.e.}, 
$\epsilon_{c,b}=100\%$). This is precisely what occurs
in the case of $s\bar s b\bar b$ final states whereas $W^\pm$
mediated SM backgrounds cannot contribute to $b\bar b b\bar b$
final states under the above assumption
(we will briefly discuss the more realistic scenario arising
from a finite efficiency for the latter in the last section).

When the sneutrino mass starts departing from $M_Z$ (or $M_W$), then
a variety of SM sub-processes could produce sizable irreducible
backgrounds, although at very heavy $M_{\tilde\nu}$ values only the
tails of the SM distributions can actually play a role. All these channels
can be conveniently grouped into general four-fermion final states,
of the type $e^+e^- e^+e^-$,
$\mu^+ \mu^- \tau^+\tau^{-}$,
$\tau^+ \tau^- \tau^+ \tau^-$,
$s \bar s s \bar s$,
$s \bar s b\bar b$ and
$b \bar b  b \bar b$, which we have generated by means of all
Feynman graphs appearing at leading order,
with the only exclusion of Higgs mediated graphs, which are
irrelevant for the first four channels, because
of the smallness of the Yukawa couplings
involved (see also footnote 6), and since they can easily be 
excluded in the last two 
cases via a suitable invariant mass cut, {\it i.e.}, $M_{b\bar b}\neq
M_{\rm Higgs}$, for any known neutral Higgs mass state of the model,
thanks to the narrowness of the Higgs boson resonances below
the $W^\pm W^{\mp *}$ threshold.
\begin{table}[!t]
\begin{center}
\begin{tabular}{|c|c|c|c|c|c|c|} \hline
$M_{\sneu}$ (GeV) &
$e^+e^- e^+e^-$ &
$\mu^+ \mu^- \tau^+\tau^{- *}$ &
$\tau^+ \tau^- \tau^+ \tau^-$ &
$s \bar s s \bar s$ &
$s \bar s b\bar b^*$ &
$b \bar b  b \bar b$ \\ \hline\hline
$100$ &0.1313 &0.3268 &0.1860 & 1.7951&1.6455  &1.0897  \\ 
      &0.8168 &0.9963 &0.9496 & 1.8394&1.6659  & 1.1262 \\ 
      &0.8168 &0.9963 &0.9496 &1.8394 & 1.6659 &1.1262  \\ 
      &206634870 & 48  & 14  & 39993 & 271 & 28  \\ \hline
$200$ &0.00269  &0.00801 &0.00401 &0.4869 &0.5385 &0.2621  \\ 
      &0.03740  &0.09613 &0.05383 & 0.7471& 0.6854& 0.4474 \\ 
      &0.06897  &0.1579  &0.09681 &0.7617 &0.6921 &0.4590  \\ 
      & 59351446  & 8.0 & 8.2 & 15451 & 163 & 16  \\ \hline
$300$ &0.000486  & 0.001532 &0.000701  &0.1459 &0.1913 &0.07331 \\ 
      &0.003192  & 0.009818 &0.004616  & 0.3039&0.3059  &0.1729 \\ 
      &0.007567  & 0.02235  &0.01083  &0.3424 &0.3262  &0.2014 \\ 
      & 23373961  & 3.7 & 1.1 & 7005  & 47  & 7.3 \\ \hline
$400$ &0.000189 & 0.0005867 &0.000282 &0.06167 &0.08462 &0.03029 \\ 
      & 0.000997& 0.003023& 0.001452 &0.1351 & 0.1432& 0.07438\\ 
      & 0.002332& 0.006881 &0.003376 & 0.1610& 0.1579& 0.09261 \\ 
      & 11211595  & 2.3 & 0.93& 2393  & 19  & 2.3 \\ \hline
$500$ &$8.476\times10^{-5}$  &0.0002556 & 0.000117&0.02832&0.03792 &0.01336 \\ 
      & 0.000226& 0.000685& 0.000320 &0.05041 & 0.05710& 0.02544\\ 
      & 0.000920& 0.002706 &0.001305& 0.07713&  0.07362& 0.04249\\ 
      & 5294773   & 0.38& 0.22& 1335  &  8  & 1.0 \\ \hline
\multicolumn{7}{|c|}
{$^*$Other c.c.'s are free from SM background}\\ \hline
\end{tabular}
\end{center}
\caption{ Same as Table \ref{mass_cuts} for $\sqrt{s}_{e^+e^-}= 1 $~TeV. }
\label{mass_cutstev}
\end{table}

The first three signatures only receive EW contributions ({\it i.e.},
they are of ${\cal O}(\alpha^4)$), whereas the last three also have
a QCD induced component of ${\cal O}(\alpha^2\alpha_s^2)$. Under 
the assumption of perfect $b$ (and $c$) quark tagging, we only need to sum
over light quark flavours in the case of  
$s \bar s s \bar s$ and $s \bar s b\bar b$ final states,
not of $b \bar b b\bar b$. However, in order to save computing time,
we have presently refrained
from doing so, as the EW and QCD contributions involving only $s$ 
(anti)quarks, as opposed to the sum over $u,d$ and $s$, already
swamp the signal in the accessible $M_{\tilde\nu}$ regions.
(All forthcoming $s \bar s s \bar s$ and $s \bar s b\bar b$  
background rates will then be reported for $s$ flavours only.)\footnote{Also,
we have used the two-loop expression for $\alpha_s$, as a function
of the energy scale $Q\equiv\sqrt{s_{\gamma\gamma}}$ and of 
$\Lambda_{\rm{QCD}}^{n_f=4}=0.230$ GeV.} 
Of course, four-fermion final states computed this way also
include the contributions
of processes of the type (\ref{Z})--(\ref{Zdecay})
and (\ref{W})--(\ref{Wdecay}).


\begin{table}[!t]
\begin{center}
\begin{tabular}{|c|c|c|c|c|c|c|} \hline
$M_{\sneu}$ (GeV) &
$e^+e^- e^+e^-$ &
$\mu^+ \mu^- \tau^+\tau^{-}$ &
$\tau^+ \tau^- \tau^+ \tau^-$ &
$s \bar s s \bar s$ &
$s \bar s b\bar b$ &
$b \bar b  b \bar b$ \\ \hline\hline
 100 & 0.00031 &  2.0 &  1.9 &  0.25 &  2.7 &  5.2\\ 
 &  0.0019 &  6.0 &  9.5 &  0.25 &  2.7 &  5.4\\ 
 &  0.0019 &  6.0 &  9.5 &  0.25 &  2.7 &  5.4\\ 
 200 &  7.6~$\times10^{-6}$ &  0.087 &  0.047 &  0.082 &  0.97 &  1.3\\ 
 &  0.00011 &  1.0 &  0.63 &  0.13 &  1.2 &  2.3 \\
 &  0.0020  & 1.7 &  1.14  &  0.13 &  1.3 & 2.3 \\
 300 &  1.2~$\times10^{-6}$ &  0.0082 &  0.0054 &  0.023 &  0.28 &  0.29\\ 
  & 8.1~$\times10^{-6}$  & 0.053  & 0.035  & 0.048  & 0.44 &  0.68\\ 
  & 1.9~$\times10^{-5}$ &  0.12  & 0.083  & 0.054  & 0.47 &  0.79\\ \hline
\end{tabular}
\end{center}
\caption{Significances $S/\sqrt B$ at $\sqrt{s_{e^+e^-}}=500$
GeV after 1 ab$^{-1}$ 
for the four-fermion final states discussed in the text,
for both signals (first three rows for MSSM set A,B and C respectively) 
and backgrounds (last row),  
for unpolarised beams, after the
cuts in (\ref{cuts}) plus the additional constraint $|M_{ff}-M_{\sneu}|<5$
GeV, for several sneutrino masses. (Here, $ff$ refer to either
lepton-lepton or jet-jet pairs and only one  $M_{\sneu}$ is required
to be reconstructed.) Notice that, for the backgrounds, no summation
over $u,d$ and $s$ (light) flavours has been performed in the case of 
signatures involving $s$ quarks.}
\label{mass_cuts_sig}
\end{table}
\begin{table}[!t]
\begin{center}
\begin{tabular}{|c|c|c|c|c|c|c|} \hline
$M_{\sneu}$ (GeV) &
$e^+e^- e^+e^-$ &
$\mu^+ \mu^- \tau^+\tau^{-}$ &
$\tau^+ \tau^- \tau^+ \tau^-$ &
$s \bar s s \bar s$ &
$s \bar s b\bar b$ &
$b \bar b  b \bar b$ \\ \hline\hline
 100 & 0.00029 & 1.5 & 1.6 & 0.28 & 3.1 &  6.5 \\
  & 0.0018 & 4.5 &  8.0 & 0.29 & 3.2 & 6.7 \\
  & 0.00179685869 & 4.54746653 &  8.0 & 0.29 & 3.2 & 6.7 \\
 200 & 1.1~$\times10^{-5}$ & 0.090 & 0.044 & 0.12 & 1.3 &  2.1 \\
  & 0.00015 & 1.1 &  0.59 & 0.19 & 1.7 & 3.5 \\
  & 0.00028 & 1.8 &   1.1  & 0.19 & 1.7 & 3.6 \\
 300&  3.2~$\times10^{-6}$ & 0.025 & 0.021 & 0.055 & 0.88 & 0.86 \\
  & 2.1~$\times10^{-5}$ & 0.16 & 0.14 &  0.11 & 1.4 & 2.0 \\
  & 4.9~$\times10^{-5}$ & 0.37 & 0.33 &  0.13 & 1.5  & 2.4 \\
 400 & 1.8~$\times10^{-6}$ & 0.012 & 0.0092 & 0.039 &  0.61 & 0.63 \\
  & 9.4~$\times10^{-6}$ & 0.063 & 0.048 & 0.087 & 1.04 & 1.6 \\
  & 2.2~$\times10^{-5}$ & 0.14  & 0.11  & 0.10  & 1.1   & 1.9 \\
 500 & 1.2~$\times10^{-6}$ & 0.013 &  0.0079 & 0.025 & 0.42 & 0.42 \\
  & 3.1~$\times10^{-6}$ & 0.035 &  0.022 &   0.044 & 0.64 &  0.80 \\
  & 1.3~$\times10^{-5}$ & 0.14  &  0.088 &   0.067 & 0.82 &  1.3 \\ \hline
\end{tabular}
\end{center}
\caption{ Same as Table \ref{mass_cuts_sig} for $\sqrt{s}_{e^+e^-}= 1 $~TeV. }
\label{mass_cutstev_sig}
\end{table}
The SM background cross-sections, 
after the cuts listed in eq.~(\ref{cuts}), are
given in Table~\ref{backgrounds}. A common feature to all rates 
is that they are basically independent of the polarisation
state of the initial particles\footnote{The huge rates for the $e^+e^- e^+e^-$
final state should not be surprising: on the one hand, because of the
$t,u$ channel soft and collinear singularities of the total cross section
(which are only regulated by the small electron mass); on the other hand, 
since we have implemented rather loose constraints
in order to avoid these, {\it i.e.}, other than
the cuts in individual energy and polar angles of eq.~(\ref{cuts}) we
only required $M_{e^+e^-}>1$ GeV on all electron-positron pairs
(this combination is sufficient to obtain a numerically stable answer).}. 
By comparing the background rates in Table \ref{backgrounds}
to those for the signals in Figures~\ref{signal_1}--\ref{signal_6},
it is clear that the former are overwhelming the latter in the
{\sl inclusive} cross sections. However, several selection cuts
can be applied in order to improve the signal-to-background ratio
($S/B$). For example, the $Z$ mediated noise in four-lepton
final states can be reduced by requiring that
no $\ell^+\ell^-$ pairs of opposite charge (with $\ell=e,\mu$)
reproduces the $Z$ mass within a few GeV (say, 3 or 4 GeV, given the
good mass resolution expected at LCs for electrons and muons). Similarly,
one can proceed for four-quark final states, by rejecting events
with one (or more) jet-jet invariant masses in the vicinity of the
$Z$ and $W^\pm$ peaks. The
$4\tau$ signature is more difficult to deal with in this respect, because
of the missing momentum carried away by the neutrinos. Finally,
QCD induced four-jet backgrounds tend to produce at least one jet-jet
pair with small invariant mass.

\begin{figure}[hbt]
\centerline{\epsfig{file=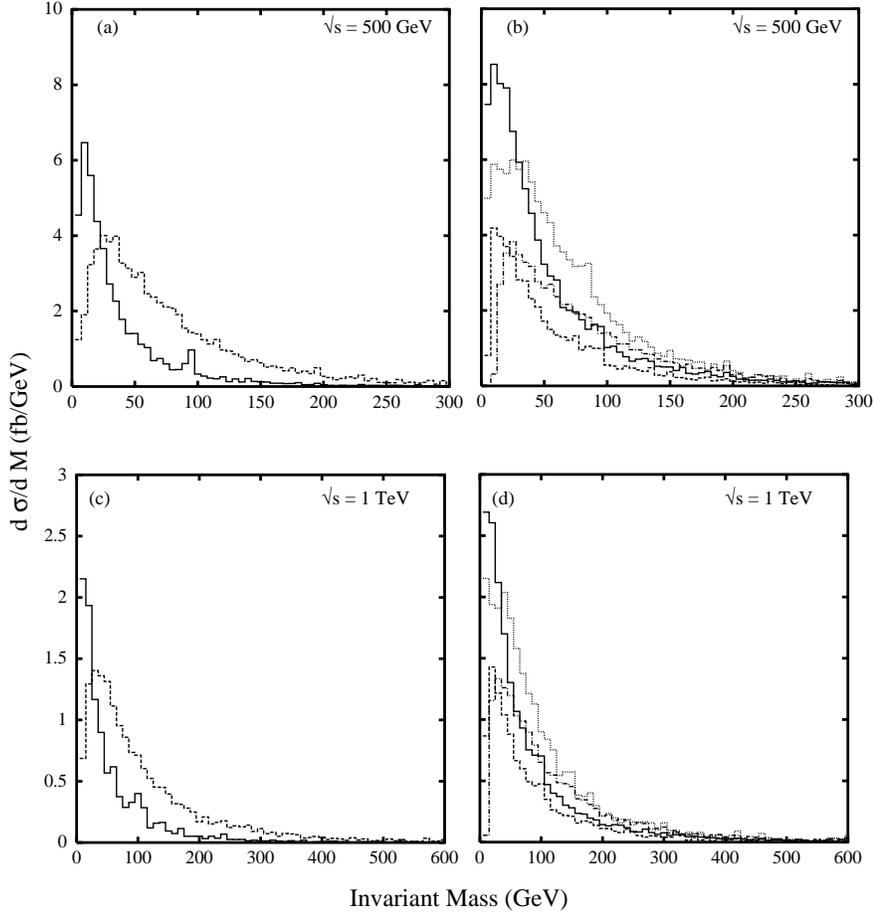,width=0.75\linewidth}}
 \vspace*{-3.5cm}
\caption{Invariant mass distributions in lepton-lepton and
jet-jet pairs reproducing the sneutrino mass in the background
processes
discussed in the text, for the case of unpolarised beams,
$\sqrt{s_{e^+e^-}}=500$ GeV and 1 TeV, after the cuts in
(\ref{cuts}) (the additional
constraint $M_{e^+e^-}>1$ GeV has been implemented for the
$e^+e^-e^+e^-$ signature, for all possible pairings with opposite
EM charge). Normalisation is to the corresponding total cross sections 
given in Table \ref{backgrounds}. All possible combinatorial combinations
have been plotted, each with the same probability, given by
the event weight divided by the number of possible pairings
in each case. Bins are 10 GeV wide. Scaling factors have to be applied,
in order to obtain the original cross sections, as follows:
(a,c)
 $\mu^+\mu^-\tau^+\tau^-$ (solid) to be scaled by 10$^{-2}$, 
 $s\bar sb\bar b $ (dashed) by 10$^{-2}$;              
(b,d)
 $e^+e^-e^+e^-$ (solid) to be scaled by 10$^{4}$, 
 $\tau^+\tau^-\tau^+\tau^-$ (dashed) by 10$^{-3}$,              
 $s\bar s s\bar s$ (dotted) no scaling,
 $b\bar bb\bar b$ (dot-dashed)  by 10$^{-3}$. 
}
\label{masses}
\end{figure}

In the very end, however, one should keep in mind that we are dealing with sneutrino masses 
that are bound by current experimental constraints to be above
the EW scale. Hence, in general, by 
restricting oneself to the most interesting mass range, sufficiently
far from the $Z$ and $W^\pm$ masses, say, 100--150 GeV or above,
the chances of extracting the RPV signals in some of the channels 
discussed become evident, if one refers to Figure~\ref{masses}
and to the rates in Figures~\ref{signal_1}--\ref{signal_6}.

By finally recalling that sneutrinos yield mass resonances that are rather
narrow (see the typical widths in the Appendix), one can further
enhance the $S/B$ by restricting the candidate
sample around the resonances. To this end, we 
present Table \ref{mass_cuts} and \ref{mass_cutstev}, where, alongside the signal
yield, the surviving background 
rates are given, after we have required that only one (di-lepton
or di-jet) invariant mass reconstructs the resonant sneutrino mass 
within 10 GeV. Notice that
we have accounted for all combinatorics in both leptonic and hadronic
final states, assuming EM charge recognition in the former but not
in the latter. Results are given for the unpolarised case, for the sake
of illustration. (In the case of polarised initial states, the pattern
is very similar.)
 
At the end of this selection, one should expect
the final states $\mu^+\mu^-\tau^+\tau^-$,
$\tau^+\tau^-\tau^+\tau^-$ and $b\bar b b\bar b$ to achieve
a significance $\sigma\equiv S/\sqrt B$ larger than 5 after 1 ab$^{-1}$ of luminosity
in the region $M_{\sneu}\sim$ 100--150 GeV, at both
$\sqrt{s_{e^+e^-}}=500$ GeV and 1 TeV, for all
MSSM parameter sets considered in the case of the hadronic signature and limitedly to set B and C 
for the leptonic ones. The overall signal rates at that luminosity 
are about 1,000 events per channel. At the higher collider
energy option, an evidence  ({\it i.e.}, $\sigma\gsim3$) of the $4b$ signal  may 
appear also in the $M_{\sneu}\sim$ 200--250 GeV interval, at least for
the MSSM sets B and C, with overall signal rates of order 500 events. 
All other signatures appear instead hopeless. 
Tables (\ref{mass_cuts_sig})--(\ref{mass_cutstev_sig}) summarise
our findings in this respect. The
typical signal would then be an excess of $\mu^+\mu^-\tau^+\tau^-$, $\tau^+\tau^-\tau^+\tau^-$
plus $b\bar b b\bar b$ 
events above the SM expectations for the corresponding four-fermion processes,
with the bulk of the events cantered in a rather narrow lepton-lepton
or jet-jet mass region corresponding to the sneutrino mass.

\section{Conclusions}

Although a full Monte Carlo simulation, including all signals
and backgrounds that we have discussed and 
in presence of both hadronisation and detector effects,
should eventually be performed
in order to put on firmer ground the results presented here,
it is clear that the latter seem rather promising at present.

In practice, if RPV couplings of the type $\lambda_{311}$,
$\lambda_{323}$, $\lp_{323}$ or $\lp_{333}$ are close 
to their current exclusion bounds, over sizable regions of the 
MSSM parameter space (particularly, for positive $\mu$ values), 
several four-fermion
signatures induced by a sneutrino, with a mass $M_{\sneu}\lsim150$ GeV if
$\sqrt{s_{e^+e^-}}=500$ GeV and $\lsim250$ GeV if $\sqrt{s_{e^+e^-}}=1$ TeV, 
produced in association with a fermion pair and decaying
itself into a second pair, can be accessed,
with the photons produced via back-scattering against the primary
electrons and positrons. The typical annual rate should be of
several hundreds to a thousand events in each of the three
channels $\mu^+\mu^-\tau^+\tau^-$,
$\tau^+\tau^-\tau^+\tau^-$ and $b\bar b b\bar b$, 
depending on the actual sneutrino mass 
and assuming a luminosity of 1 ab$^{-1}$. 

Furthermore, since typical SM backgrounds have been seen to be less
sensitive than the signals to the polarisation state of the incoming particles,
one may exploit the latter in order to improve the discovery potential of
RPV signals at future LCs. If a high, but not unrealistic, 
degree of polarisation of both laser photons and leptonic beams 
can be achieved, this can be exploited to push the discovery
reach in sneutrino mass somewhat beyond the mentioned $M_{\sneu}$ values
in the $\tau^+\tau^-\tau^+\tau^-$ and $b\bar b b\bar b$ final states
({\it i.e.}, those most massive in the leptonic and hadronic case, respectively)
at 500 GeV. In fact, at this energy, the polarisation combination in which the electron and
positron helicities have the opposite sign and are also opposite to those 
of the corresponding laser photons, {\it i.e.}, $(+-)$, yields, for sneutrino
masses in the 100--250 GeV region,  signal rates somewhat higher than 
those induced in the other cases (including that of unpolarised beams), 
up to a factor of 2. In contrast, for the
$\mu^+\mu^-\tau^+\tau^-$ final state, it is the unpolarised configuration,
{\it i.e.}, (00),
the most suitable for sneutrino searches in the above mass range. Once
the collider energy is raised to 1 TeV, differences between the three polarisation
combinations almost disappear if $M_{\tilde\nu}\lsim250$ GeV. The
polarisation state  in which the electron and
positron helicities have the same sign and opposite to the one 
of the laser photons, {\it i.e.}, (++), would turn out extremely
useful for heavies sneutrino masses, say, when $M_{\tilde\nu}\gsim\sqrt{s_{e^+e^-}}/2$,
as  here  signal rates are consistently and significantly above
those induced by the other polarisations, up to a factor of 4 in some instances.
Unfortunately, this mass interval is unattainable through present LC designs
(TESLA, NLC and JLC) and will have to attend for higher collider energies,
such as those foreseen for CLIC (${\sqrt s_{e^+e^-}}\gsim 3$ TeV). 

Before closing, two final considerations are in order.
Firstly, recall that, as a bonus of the
production process considered here, some leptonic signatures which
are flavour changing, such as $\mu^+\mu^+\tau^-\tau^-$ and
$\mu^-\mu^-\tau^+\tau^+$, would come practically free from
SM background, hence promptly detectable at a future LC. 
Secondly, that we have not included the effect of finite
experimental efficiency in tagging leptons and jets,
so that our final significances
may be somewhat over-estimated. The comment particularly applies to
the $4b$ final state, for which we have assumed throughout a 100\% efficiency
for a quadruple $b$-tagging. If one adopt a more realistic 70\%
per $b$-jet, significances in the last columns 
 in Tables (\ref{mass_cuts_sig})--(\ref{mass_cutstev_sig})
would go down by a factor of 2, hampering seriously the scope of the
hadronic channel. However, one may alternatively consider to tag
only a subset of the four $b$-quarks. The correct estimate of
the potential in this channel clearly depends on the tagging strategy
of $b$-quarks and can only be obtained in the context of 
a fully hadronic environment and in presence of detector
effects, which were laking in our study. However,
in the worse case scenario in which the $b\bar bb\bar b$ signature 
of RPV sneutrinos produced in associated mode is swamped
by the backgrounds, one should still be able to resort to
$\mu^+\mu^-\tau^+\tau^-$ and $\tau^+\tau^-\tau^+\tau^-$.
Altogether, we consider the subject raised in this paper of 
relevance for the physics of future LCs and look forward
to experimental studies in the context of the current LC
workshops.

\section{Acknowledgements}
We thank Abdesselam Arhrib for discussions. 
The work of DKG is supported by the  National Science
Council of Taiwan under the grant  NSC 90-2811-M-002-054 and
from the Ministry of Education Academic Excellence Project 89-N-FA01-1-4-3 of
Taiwan. 
\newpage
\section{Appendix}
As intimated in the main text, we reproduce here the sneutrino partial
widths in the two-body decay channels considered in the paper, 
see Table~\ref{widths}, for the MSSM
parameter sets given in Table~\ref{set_mssm}.
\begin{table}[ht]
\begin{center}
\begin{tabular}{|c|c|c|c|c|c|c|} \hline
$M_{\tilde \nu}$~(GeV) & $e^+e^-$ & $\mu^+\tau^-$ 
& $\tau^+\tau^-$ & $s \bar{s} $ 
& $s \bar {b}$ & $ b \bar {b}$\\
\hline
100  & 0.0238 & 0.0297 & 0.0248 & 0.8269 & 1.6270& 0.6169 \\
\hline
200 & 0.6474& 0.6593 & 0.6495& 2.2536& 3.8640& 1.8447\\
\hline
400 &4.1109& 4.1345& 4.1151 & 7.3232 & 10.5492 & 6.5109\\
\hline
\end{tabular}
\end{center}
\begin{center}
\begin{tabular}{|c|c|c|c|c|c|c|} \hline
$M_{\tilde \nu}$~(GeV) & $e^+e^-$ & $\mu^+\tau^-$ 
& $\tau^+\tau^-$ & $s \bar{s} $ 
& $s \bar {b}$ & $ b \bar {b} $\\
\hline
100  & 0.0038 & 0.0097 & 0.0049 & 0.8069 & 1.6070& 0.5970 \\
\hline
200 & 0.0642& 0.0760 & 0.0663& 1.6704& 3.2808& 1.2614\\
\hline
400 &0.8613&0.8850&0.8655 & 4.0737 & 7.2996 & 3.2613\\
\hline
\end{tabular}
\end{center}
\begin{center}
\begin{tabular}{|c|c|c|c|c|c|c|} \hline
$M_{\tilde \nu}$~(GeV) & $e^+e^-$ & $\mu^+\tau^-$ 
& $\tau^+\tau^-$ & $s \bar{s} $ 
& $s \bar {b} $ & $ b \bar {b}  $\\
\hline
100  & 0.0038 & 0.0097 & 0.0049 & 0.8069 & 1.6070& 0.5970 \\
\hline
200 & 0.0294& 0.0413 & 0.0315& 1.6356& 3.2460& 1.2267\\
\hline
400 &0.3975&0.4212&0.4017 & 3.6099 & 6.8359 & 2.7976\\
\hline
\end{tabular}
\end{center}
\caption{Sneutrino decay width (in GeV) in the RPV decay
channels relevant to our analysis for the MSSM parameter sets
A (top), B (middle) and C (bottom).}  
\label{widths}
\end{table}
\newpage
\def\pr#1, #2 #3 { {\em Phys. Rev.}         {\bf #1},  #2 (19#3)}
\def\prd#1, #2 #3{ {\em Phys. Rev.}        {D \bf #1}, #2 (19#3)}
\def\pprd#1, #2 #3{ {\em Phys. Rev.}       {D \bf #1}, #2 (20#3)}
\def\prl#1, #2 #3{ {\em Phys. Rev. Lett.}   {\bf #1},  #2 (19#3)}
\def\pprl#1, #2 #3{ {\em Phys. Rev. Lett.}   {\bf #1},  #2 (20#3)}
\def\plb#1, #2 #3{ {\em Phys. Lett.}        {\bf B#1}, #2 (19#3)}
\def\pplb#1, #2 #3{ {\em Phys. Lett.}        {\bf B#1}, #2 (20#3)}
\def\npb#1, #2 #3{ {\em Nucl. Phys.}        {\bf B#1}, #2 (19#3)}
\def\nnpb#1, #2 #3{ {\em Nucl. Phys.}        {\bf B#1}, #2 (20#3)}
\def\prp#1, #2 #3{ {\em Phys. Rep.}        {\bf #1},  #2 (19#3)}
\def\zpc#1, #2 #3{ {\em Z. Phys.}           {\bf C#1}, #2 (19#3)}
\def\epj#1, #2 #3{ {\em Eur. Phys. J.}      {\bf C#1}, #2 (19#3)}
\def\mpl#1, #2 #3{ {\em Mod. Phys. Lett.}   {\bf A#1}, #2 (19#3)}
\def\ijmp#1, #2 #3{{\em Int. J. Mod. Phys.} {\bf A#1}, #2 (19#3)}
\def\ptp#1, #2 #3{ {\em Prog. Theor. Phys.} {\bf #1},  #2 (19#3)}
\def\jhep#1, #2 #3{ {\em J. High Energy Phys.} {\bf #1}, #2 (19#3)}
\def\pjhep#1, #2 #3{ {\em J. High Energy Phys.} {\bf #1}, #2 (20#3)}
\def\epj#1, #2 #3{ {\em Eur. Phys. J.}        {\bf C#1}, #2 (19#3)}
\def\eepj#1, #2 #3{ {\em Eur. Phys. J.}        {\bf C#1}, #2 (20#3)}
\def\mmpl#1, #2 #3{ {\em Mod. Phys. Lett.}   {\bf A#1}, #2 (20#3)}

\newpage
\newpage
\newpage
\newpage
\newpage
\newpage
\newpage
\newpage
\newpage
\newpage
\end{document}